\documentclass[letterpaper,twocolumn,10pt]{article}
\usepackage{usenix-2020-09}

\newcommand{\thiswork}{\textsf{Syncopate}}

\newcommand{\Fig}[1]{Fig.~{#1}}
\newcommand{\Tbl}[1]{Tbl.~{#1}}

\usepackage{graphicx} 
\usepackage[dvipsnames,table,xcdraw]{xcolor}
\usepackage{amsmath}

\usepackage{lipsum}  

\usepackage{makecell}

\usepackage{booktabs}

\usepackage{array}% for m

\usepackage{makecell}

\usepackage[most]{tcolorbox}
\usepackage{amssymb}
\definecolor{darkblue}{HTML}{ae230d}
\definecolor{darkred}{HTML}{b1001d}

\usepackage{circledtext}
\newcommand{\ballnumber}[1]{\circledtext*[charf=\Large]{#1}}

\usepackage{tikz}
\newcommand{\circlednumber}[3][white]{%
  \tikz[baseline=(char.base)]{
    \node[
    shape=circle, 
    fill=#2, 
    text=#1, 
    draw=#1,
    inner sep=0.6pt] (char) {#3};
  }%
}

\renewcommand{\paragraph}[1]{\vspace{0.1cm}\noindent\textbf{#1}}

\usepackage{multirow}

\usepackage{pifont}
\newcommand{\cmark}[1][black]{\textcolor{#1}{\ding{52}}}
\newcommand{\xmark}[1][black]{\textcolor{#1}{\ding{55}}}

\usepackage{color}
\definecolor{codegreen}{rgb}{0,0.6,0}
\definecolor{codegray}{rgb}{0.5,0.5,0.5}
\definecolor{codepurple}{rgb}{0.58,0,0.82}
\definecolor{backcolour}{rgb}{0.95,0.95,0.92}
\definecolor{textblue}{rgb}{.2,.2,.7}
\definecolor{textred}{rgb}{0.54,0,0}
\definecolor{textgreen}{rgb}{0,0.43,0}
\definecolor{codered}{rgb}{201,72,12}

\usepackage[T1]{fontenc}
\usepackage[scaled=0.85]{beramono} 
\usepackage{listings}
\usepackage{multirow}
\usepackage{setspace}

\newtoggle{usecomment}
\settoggle{usecomment}{true}

\usepackage[available]{usenixbadges}

\newcommand{\rev}[1]{{#1}}
\begin{document}
%-------------------------------------------------------------------------------
\pagestyle{empty}
%don't want date printed
\date{}

% \title{\Large \bf \thiswork{}: Automatic Fine-Grained Compute-Communication\\ Overlap via  Chunk-Centric Scheduling}
\title{\Large \bf \thiswork{}: Efficient Multi-GPU AI Kernels via Automatic Chunk-Centric Compute-Communication Overlap}

\author{
{\rm Xinwei Qiang$^{1,*}$, Yue Guan$^{1,*}$, Zhengding Hu$^1$, Keren Zhou$^{3,4}$, Yufei Ding$^{1,2}$, Adnan Aziz$^2$}\\
$^1$University of California, San Diego, $^2$Meta, 
$^3$George Mason University, 
$^4$OpenAI\\
$^1$\{x1qiang, yueguan, zhh068, yufeiding\}@ucsd.edu\\
$^2$\{adnanaziz\}@meta.com\\
$^3$\{kzhou6\}@gmu.edu
}

\maketitle
% {\let\thefootnote\relax\footnote{{$^*$Equal contribution.}}}
% {\let\thefootnote\relax\footnote{{$^*$yueguan@ucsd.edu}}}

\begin{abstract}
Communication has become a first-order bottleneck in large-scale GPU workloads, and existing distributed compilers address it mainly by overlapping whole compute and communication kernels at the stream level. This coarse granularity incurs extra kernel launches, forces device-wide synchronizations at kernel boundaries, and leaves substantial slack when the slowest tile or kernel stretches the communication tail. We present \thiswork{}, a compiler and runtime that enable automatic fine-grained overlap around a single fused compute kernel. \thiswork{} introduces a communication chunk abstraction that decouples communication granularity from kernel structure and backend mechanisms, allowing chunk-level plans to be ported from existing distributed compilers, written directly by users, or instantiated from reusable templates. Given a local Triton kernel and a chunk schedule, \thiswork{} performs transformations to align computation with chunk availability. Implemented as a source-to-source compiler on Triton, \thiswork{} delivers an average end-to-end speedup of 1.3$\times$ and up to 4.7$\times$ on multi-GPU workloads. 
Our code is open-sourced at \url{https://github.com/tie-pilot-qxw/syncopate}.

% \thiswork{} introduces a communication \emph{chunk} abstraction that decouples high-level communication schedules from low-level implementations, allowing chunk-level plans to be ported from existing distributed compilers, written directly by users, or instantiated from reusable templates. Given an annotated local Triton kernel and a chunk schedule, \thiswork{} builds a unified chunk--tile dependence IR, rewrites the tile scheduler to align computation with chunk availability, and selects among copy-engine, TMA, and load/store backends through inter- and intra-chunk autotuning. Implemented as a source-to-source compiler on Triton and integrated with PyTorch distributed, \thiswork{} generates efficient fused distributed kernels that deliver substantial end-to-end speedups over state-of-the-art kernel-level overlap across a range of training and inference workloads.
\end{abstract}

{\let\thefootnote\relax\footnote{{$^*$Xinwei and Yue contributed equally.}}}

\section{Introduction} \label{sec:introduction}

Communication has become a first-order bottleneck for training and serving large neural networks on multi-GPU systems. Even with high-bandwidth interconnects such as NVLink~\cite{nvidia_nvlink_whitepaper} and NVSwitch, collective operations like AllGather, ReduceScatter, and All-to-All frequently dominate end-to-end latency for tensor-parallel feed-forward layers and attention layers. To hide this cost, recent systems and distributed compilers aggressively search for schedules that overlap computation and communication at the kernel level. Given a computation graph and a device mesh, these compilers select parallelization strategies, insert communication kernels, and assign compute and communication kernels to streams so that multiple kernels can run concurrently. This kernel-level overlap~\cite{wang2024dominoeliminatingcommunicationllm, chen2024centauri, shoeybi2019megatron, zhu2025nanoflowoptimallargelanguage, gond2025tokenweaveefficientcomputecommunicationoverlap, OverlapCommunicationwithDependentComputationviaDecomposition} has become the default mechanism for improving utilization in distributed settings.

\begin{figure*}
    \centering
    \includegraphics[width=\linewidth]{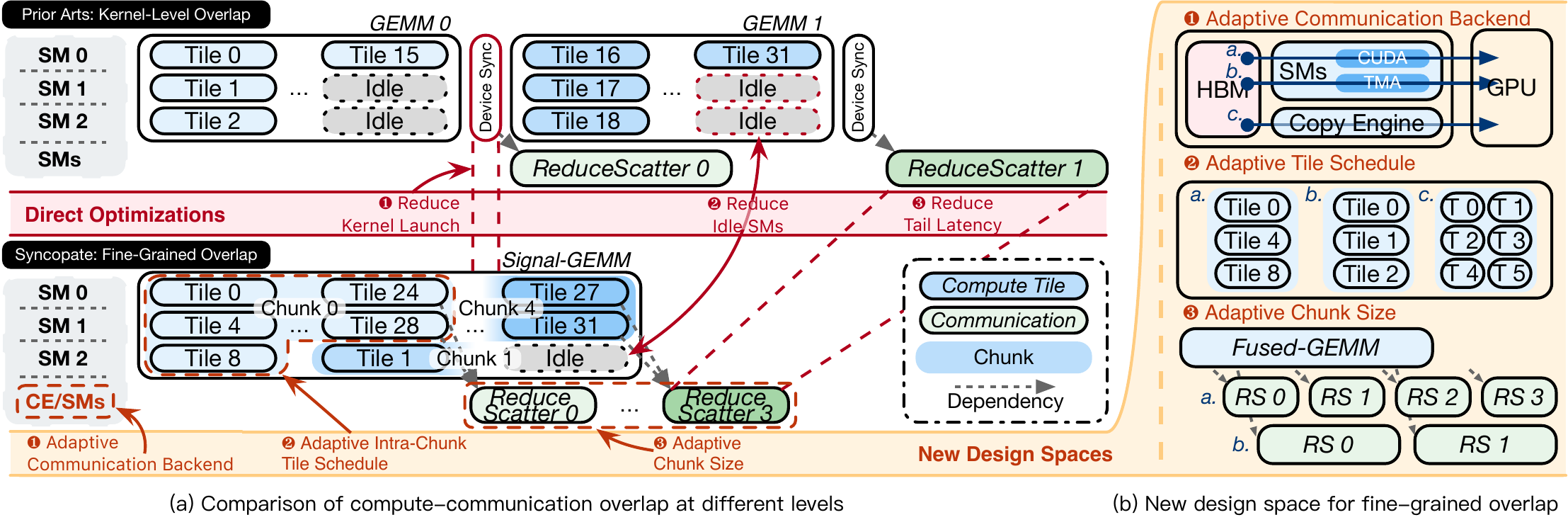}
    \vspace{-2em}
    \caption{Motivating example of \thiswork{}. \textcolor{darkred}{Red numbers}  show the direct improvements gained by fine-grained overlap over kernel-level overlap, while \textcolor{darkblue}{orange numbers}  show the additional improvements from the new design space enabled by \thiswork{}.}
    \label{fig:motivating_example}
    \vspace{-1em}
\end{figure*}

However, kernel-level overlap is fundamentally insufficient for fully hiding communication latency. As illustrated in \Fig{\ref{fig:motivating_example}}, this kernel-level scheduling forces a device-wide synchronization at every kernel boundary and incurs extra launch and sync overhead for each communication phase ({\small \circlednumber{darkred}{1}}). Moreover, by splitting computation into multiple shorter kernels, the work within each launch is further partitioned into waves of compute tiles on each SM, increasing the fraction of time SMs sit idle; even when part of the data needed by later kernels is already available, tiles in the current wave must wait for the slowest tile to finish ({\small \circlednumber{darkred}{2}}). Finally, the coarse granularity of kernel-level overlap leaves a long segment of communication at the end of the timeline that receives little or no overlap ({\small \circlednumber{darkred}{3}}).

This motivates us to overlap computation and communication at a finer intra-kernel granularity. Such fine-grained overlap opens up new design space for improving end-to-end efficiency. As shown by the yellow region in \Fig{\ref{fig:motivating_example}}, \thiswork{} launches communication directly from within the fused kernel, rather than delegating to external communication libraries, giving the compiler explicit control over which hardware backend to use for each transfer (copy engine, tensor memory accelerator, or load/store on CUDA cores) as shown with {\small \circlednumber{darkblue}{1}}. This also allows us to explore much smaller communication granularities without incurring additional kernel-launch overhead, and to tune the chunk size that best balances link throughput against synchronization cost ({\small \circlednumber{darkblue}{2}}). Finally, because tiles and communication are orchestrated inside a single kernel, we can reshape the intra-kernel tile schedule to track communication progress while still preserving locality in the register, shared-memory, and cache hierarchy ({\small \circlednumber{darkblue}{3}}).

We present \thiswork{}, a compiler that turns the vision of fine-grained overlap for distributed kernel generation into a practical and general system. At the heart of \thiswork{} is the notion of a communication chunk, which represents a logical block of data associated with a particular communication operation and the tiles that produce or consume it. This abstraction is motivated by a key observation: fine-grained overlap requires a communication granularity that flexibly matches how tiles generate data and how communication backends consume it, rather than assuming that tile-level or full-kernel granularity is always appropriate. By making chunks explicit, \thiswork{} can represent a wide range of overlap patterns and enable the compiler to reason about when each chunk becomes available, which tiles produce or use it, and how these dependencies interact with fused multi-stage tensor programs and potentially irregular collectives. This abstraction exposes a small set of principled knobs, including chunk size, backend choice, and tile order, that define the design space later explored by our autotuner.

Building on this abstraction, \thiswork{} implements a source-to-source compiler and runtime that transform standard Triton~\cite{triton} kernels and a high-level chunk-based communication plan into fused distributed kernels capable of fine-grained compute–communication overlap. The compiler restructures the kernel’s tile execution to follow communication progress and selects appropriate backends to realize each chunk transfer, while a lightweight runtime executes these transfers and integrates seamlessly with PyTorch distributed~\cite{pytorch2} so that \thiswork{} kernels can replace standard operators with only minimal changes to user code. \thiswork{} further performs inter- and intra-chunk autotuning, adjusting chunk sizes, backend choices, SM allocations, and tile schedules, to consistently achieve high performance across diverse operators. This implementation strategy makes the abstract chunk-based model concrete, enabling rapid prototyping of new overlap policies while remaining compatible with real distributed workloads and production communication stacks.

In summary, this work makes the following contributions:
\begin{itemize}
    \item We introduce a chunk-based abstraction that decouples high-level overlap intent from low-level implementation, enabling fine-grained compute-communication overlap inside distributed kernels.
    \item We design and implement a compiler pipeline and runtime that takes annotated local kernels and chunk-level communication plans, then generates efficient fused distributed kernels with inter- and intra-chunk autotuning.
    \item We evaluate \thiswork{} on a diverse set of distributed workloads and observe an average speedup of 1.3$\times$ on common operators, with improvements reaching up to 4.7$\times$ in the best cases.
\end{itemize}

\section{Background and Related Works} \label{sec:background}

\subsection{Distributed Compilers}

As model sizes grew, tensor compilers\cite{chen2018tvm, autotvm, feng2023tensorir, ansor, shao2022tensor, wu2025mirage} incorporated basic multi-GPU support\cite{jangda2022breaking, santhanam2021distir, alabed2025partir, PrimePar, xu2021gspmdgeneralscalableparallelization, shazeer2018meshtensorflowdeeplearningsupercomputers}, typically by composing single-GPU kernels with predefined collective primitives. Systems like Alpa\cite{zheng2022alpa}, Mercury\cite{mercury}, and other recent distributed compilers extend this idea by searching over communication patterns and schedules at the level of whole kernels: given a parallelization strategy, they first construct a communication plan containing AllGather, ReduceScatter, or All-to-All operators, and then explore different mappings of compute and communication kernels to streams in order to maximize kernel-level overlap (Table~\ref{tab:related_work}). This design has made distributed training far more accessible, but it also bakes in a rigid abstraction boundary: communication is planned as a sequence of full-kernel collectives whose launch and completion times are the basic units of scheduling. As a result, all these distributed compilers focus their search on the communication plan at the kernel level and are fundamentally blind to finer-grained opportunities inside kernels, such as overlapping per-shard or per-tile communication with computation, reusing remote data across tiles, or exploiting topology-aware pipelining within a fused kernel. Once a kernel is chosen and its associated collective is placed, the compiler can only treat it as an atomic black box, leaving significant intra-kernel overlap potential untapped even under an ``optimal'' kernel-level schedule.

\begin{table}
\centering
\caption{Comparison of projects on distributed operations.}
\label{tab:related_work}
\resizebox{\linewidth}{!}{
\begin{tabular}{lccccc}
\toprule
Projects          & Granularity & Compute & Communication & Schedule & Performance \\ \midrule
\rowcolor[HTML]{EFEFEF} 
\multicolumn{6}{c}{Automatic Approaches}                                           \\
Alpa\cite{zheng2022alpa}          & Kernel      & Auto    & Auto          & Template &   \cmark{}          \\
\rowcolor[HTML]{EFEFEF} 
Mercury\cite{mercury}           & Kernel      & Auto    & Auto          & Auto     &   \cmark{}\cmark{}          \\ \midrule
\multicolumn{6}{c}{Manual Implementations}                                         \\
\rowcolor[HTML]{EFEFEF} 
Flux\cite{chang2024flux}              & Tile       & Manual  & Manual        & Manual   &     \cmark{}\cmark{}        \\
AsyncTP\cite{OverlapCommunicationwithDependentComputationviaDecomposition}           & Tile        & Manual  & Manual        & Manual   &    \cmark{}\cmark{}        \\
\rowcolor[HTML]{EFEFEF} 
FlashOverlap\cite{hong2025flashoverlap}      & Chunk        & Manual  & Manual        & Manual   &    \cmark{}\cmark{}\cmark{}          \\
\midrule
\multicolumn{6}{c}{Domain Specific Languages}                                         \\
\rowcolor[HTML]{EFEFEF}
ThunderKittens~\cite{thunderkitten, parallelkitten}        & Tile        & Manual  & Manual        & Manual   &    \cmark{}\cmark{}\cmark{}          \\ 
TritonDistributed\cite{zheng2025triton, zheng2025tilelink} & Chunk        & Manual  & Manual        & Manual   &    \cmark{}\cmark{}\cmark{}          \\
\midrule
\rowcolor[HTML]{EFEFEF}
\textbf{\thiswork{}}       & Chunk        & Auto    & Auto          & Template &     \cmark{}\cmark{}\cmark{}         \\ \bottomrule
\end{tabular}
}
\end{table}

\subsection{\rev{Manual} Kernel Designs}

A complementary line of work pushes beyond kernel-level overlap and instead hand-crafts fine-grained pipelines that interleave communication and computation at the level of tiles, tokens, or shards. Flux\cite{chang2024flux} fuses GEMM with collectives at tile granularity and over-decomposes work to maximize overlap for both training and inference, while Comet\cite{zhang2025comet} targets MoE with a shared-tensor abstraction and NVSHMEM-backed buffers to overlap token-wise communication with tile-wise compute at production scale. FlashOverlap\cite{hong2025flashoverlap} uses lightweight readiness signaling and layout reordering to trigger overlap with standard NCCL collectives without modifying existing compute kernels, and systems like TritonDistributed\cite{zheng2025triton, zheng2025tilelink} introduce tile-centric or OpenSHMEM-style primitives to Triton~\cite{triton} that make it easier to author overlapped kernels (e.g., fused AllGather/GEMM or GEMM/ReduceScatter) in a domain-specific language.

Despite these advances, all of these systems fundamentally rely on manual, operator-specific engineering: experts must design fused kernels, choose tiling and buffering schemes, and reason about synchronization and communication protocols for each new model architecture or hardware platform. The resulting implementations are highly optimized but difficult to generalize or retarget, and they do not provide a general compiler abstraction for expressing and reusing fine-grained overlap patterns across operators. Emerging fine-grained DSLs and primitives greatly lower the barrier to writing overlapped kernels, but they still place the burden of discovering effective overlap strategies, encoding dependency structure, and validating correctness squarely on the programmer. \rev{In practice, users still need to write or modify DSL-level communication and computation code, explicitly manage signal/wait, buffer ownership, and memory ordering, and revalidate these low-level details for each operator and communication pattern.}

\subsection{Communication Backends}

\begin{table}[]
\caption{Comparison of various GPU communication mechanisms.}
\label{tab:comm_backend}
\resizebox{\linewidth}{!}{
\begin{tabular}{lclcc}
\toprule
            & Hardware    & Programming               & Collective & Bandwidth \\
\midrule
\rowcolor[HTML]{EFEFEF}
Copy Engine & Copy Engine & Host Launch               &   \xmark[red]         & \cmark{}\cmark{}\cmark{}         \\
TMA         & SM          & Async. Instruction &    \xmark[red]        &   \cmark{}\cmark{}        \\
\rowcolor[HTML]{EFEFEF}
Load/Store  & SM          & Sync.  Instruction             &    \cmark[green]        & \cmark{}  \\
\bottomrule       
\end{tabular}
}
\end{table}

% The copy engine is good because it has the best peak bandwidth and don't utilize the computation SM resources. But it requires host-launch kernels and reuiqres a larger trasfer data size (8MB) to achieve high bandwidth.
% The tma has good bandwidth and using dedicated asynchrounous tensor memory accelerator. But it requires SM to launch the instruction and does not support inter-node or collective reduction. But it only requires a small fraction amount of SMs to reach peak bandwidth.
% The load and store instruction with local register has slightly lower peak bandwidth yet support collective reduction communication on the switch. But it reuiqres SM resources and is synchrounous.
\begin{figure*}
    \centering
    \includegraphics[width=0.25\linewidth]{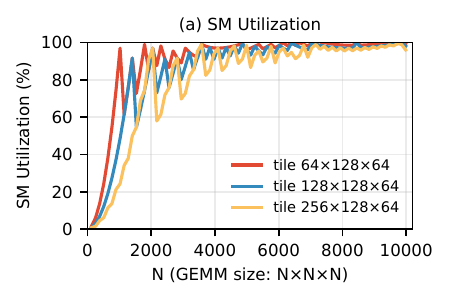}~
    \includegraphics[width=0.25\linewidth]{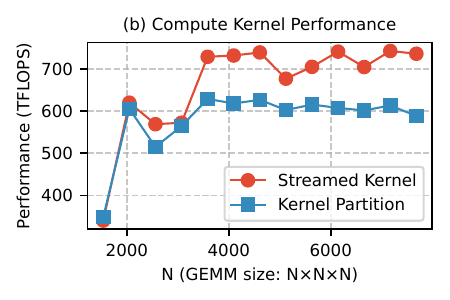}~
    \includegraphics[width=0.5\linewidth]{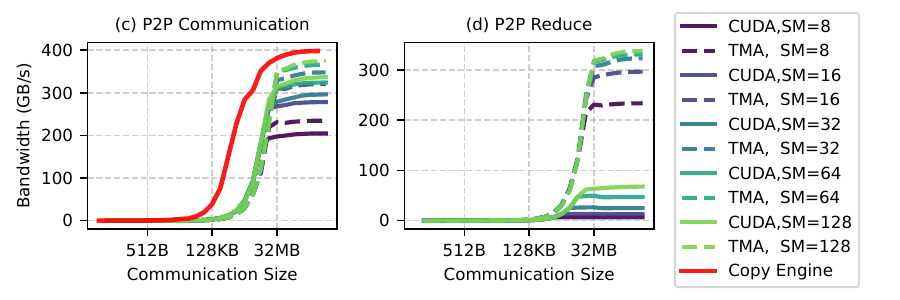}
    \caption{Motivation experiment results. (a) SM utilization under different GEMM sizes and tile sizes. (b) Performance comparison between a streamed GEMM kernel and a kernel-partitioned baseline. (c,d) Bandwidth of different communication backends under varying message sizes.}
    \label{fig:motivation}
\end{figure*}

Modern GPU systems~\cite{NVIDIA_Hopper} expose several mechanisms for moving tensors across devices, each with distinct performance and programmability trade-offs (\Tbl{\ref{tab:comm_backend}}). Copy Engines saturate NVLink~\cite{nvidia_nvlink_whitepaper} at 400 GB/s per direction on H100 and run independently of the SMs. Therefore, they do not consume compute resources and are ideal when communication can be decoupled from computation. However, they are typically driven by host APIs and can only transfer contiguous data, so high-dimensional strided tensors must be decomposed into many smaller transfers, each requiring a separate launch costing around 2-3$\mu$s, which can significantly reduce the effective bandwidth as each transfer time is also very short.

Tensor Memory Accelerator (TMA)~\cite{NVIDIA_Hopper} paths achieve high bandwidth using dedicated asynchronous tensor-copy hardware, and can achieve a throughput of 300+ GB/s with only about 16 SMs issuing TMA instructions. This makes TMA attractive for overlapping structured tensor movement with computation within a node. At the same time, TMA must be launched by SM threads and does not currently support inter-node communication or in-network (switch-based) collective reduction, which limits its applicability to intra-node, point-to-point patterns.

Finally, plain load/store-based communication, often combined with registers and shared memory, attains slightly lower peak bandwidth than copy engines or TMA~\cite{parallelkitten} but is significantly more flexible. These mechanisms integrate naturally with switch-based collective reduction (NVSHARP~\cite{nvidia_sharp_v300}), enabling fine-grained per-shard communication and in-network reductions. The downside is that they consume SM resources and are synchronous from the issuing warp's perspective, so they are harder to pipeline and require careful scheduling to hide communication latency.

\section{Motivation} \label{sec:motivation}

To fully exploit modern GPUs, distributed training systems must overlap computation and communication. Prior work mainly relies on coarse, kernel-level partitioning of computation kernels, which leaves substantial performance on the table. In this section, we revisit the motivating example in Figure~\ref{fig:motivating_example} using detailed microbenchmarks (Figure~\ref{fig:motivation}), and show three key insights:

% : (1) Kernel-level overlap is fundamentally limited by SM under-utilization and kernel-launch overheads. (2) Communication efficiency varies sharply with transfer granularity and backend selection. (3) Effective intra-kernel overlap therefore requires a communication unit with tunable granularity and a stable interface across communication backends.

% (1) kernel-level overlap is fundamentally limited by SM under-utilization and kernel-launch overheads, and (2) fine-grained, intra-kernel overlap unlocks new design spaces in communication backends, tile schedules, and chunk sizes that are difficult to explore manually.

\begin{tcolorbox}[colframe=black, colback=white, coltitle=black, boxrule=0.5mm, left=0.5em, right=0.5em, top=0.5em, bottom=0.5em]
\ding{72}\emph{\textbf{Insight 1: Limitations of Kernel-Level Overlap.}  Kernel-level overlap is fundamentally limited by SM under-utilization and kernel-launch overheads.}
\end{tcolorbox}

% \paragraph{Limitations of Kernel-Level Overlap.}
Figure~\ref{fig:motivation}(a) reports SM utilization for different GEMM sizes under several commonly used tile-size configurations. Large GEMMs provide enough tile waves to saturate the SMs across all configurations. As the GEMM size decreases, fewer tile waves are generated, and the partially filled last wave dominates a larger portion of execution, causing SM utilization to drop due to wave quantization. Partitioning a GEMM into many sub-kernels forces each launch to operate on a smaller shape, pushing execution into exactly this low-utilization regime. This effect directly limits the benefit of kernel-level partition-based overlap (corresponding to \ballnumber{2} in Figure~\ref{fig:motivating_example}), since overlapping many small kernels simply wastes SM capacity.

% However, as the GEMM size decreases, the baseline kernel-partitioned design used for overlap suffers from a sharp drop in utilization, because each sub-kernel has too little work to keep all SMs busy and cannot fully amortize scheduling overheads. 

Figure~\ref{fig:motivation}(b) compares the end-to-end performance of (i) a baseline that partitions GEMM into multiple small kernels for overlap and (ii) a streamed GEMM kernel that internally pipelines tiles. Although both variants execute the same arithmetic operations, the kernel-partitioned baseline incurs substantial performance loss due to extra kernel launches (\ballnumber{1}) and the SM under-utilization observed in Figure~\ref{fig:motivation}(a). In contrast, the streamed kernel maintains high utilization by exposing fine-grained compute tiles within a single launch, enabling overlap without fragmenting the workload. These results show that simply launching more kernels is not an effective path toward overlap; we need mechanisms that expose intra-kernel concurrency while preserving efficient GPU execution.

\begin{tcolorbox}[colframe=black, colback=white, coltitle=black, boxrule=0.5mm, left=0.5em, right=0.5em, top=0.5em, bottom=0.5em]
\ding{72}\emph{\textbf{Insight 2: Granularity and Backend Effects.} Communication efficiency varies sharply with transfer granularity and backend selection.}
\end{tcolorbox}

% \paragraph{Granularity and Backend Effects.}
Communication efficiency varies sharply with transfer granularity and backend behavior. Figures~\ref{fig:motivation}(c) and~\ref{fig:motivation}(d) show the achieved bandwidth of different communication backends as we vary the transfer size and the number of SMs. Each backend shows distinct scaling behavior: some reach peak bandwidth at moderate transfer sizes, while others require larger transfers or more SMs to reach their full potential. Moreover, different backends support different communication patterns, such as point-to-point transfers versus reductions.

Taken together, these granularity and backend effects imply that the optimal configuration depends jointly on (i) the compute tile size, which determines the cadence at which results become available, (ii) the communication transfer size, which trades off latency and bandwidth, and (iii) the choice of communication backend. Coarse-grained kernel-level overlap cannot flexibly coordinate these parameters, since compute and communication are implemented as separate kernels with rigid interfaces. Intra-kernel overlap, by coordinating computation and communication at the same time, can align tile production with backend-specific sweet spots and select transfer granularities that sustain high utilization across SMs and copy engines.

% Fine-grained overlap also creates new opportunities to jointly design computation and communication at the intra-kernel level. Figures~\ref{fig:motivation}(c) and~\ref{fig:motivation}(d) show the achieved bandwidth of different communication backends (e.g., CUDA, TMA, and copy engines) as we vary the message size and the number of SMs. Each backend exhibits distinct scaling behavior: some backends saturate quickly with moderate chunk sizes but offer limited peak bandwidth, while others require larger chunks or more SMs to reach their full potential. Moreover, different backends support different communication patterns, such as point-to-point transfers versus reductions.

% In the context of our motivating example, these observations imply that the optimal configuration depends jointly on (i) the compute tile size, which determines the cadence at which results become available, (ii) the communication chunk size, which trades off latency and bandwidth, and (iii) the choice of communication backend. Coarse-grained kernel-level overlap cannot flexibly coordinate these parameters, since compute and communication are implemented as separate kernels with rigid interfaces. By contrast, fine-grained, intra-kernel overlap, where compute tiles and communication chunks are orchestrated within a single pipeline, enables us to match tile boundaries to backend-specific sweet spots and dynamically adjust chunk sizes to maintain high utilization across SMs and copy engines.

\begin{tcolorbox}[colframe=black, colback=white, coltitle=black, boxrule=0.5mm, left=0.5em, right=0.5em, top=0.5em, bottom=0.5em]
\ding{72}\emph{\textbf{Insight 3: A Unified Unit for Intra-Kernel Overlap.} Effective intra-kernel overlap therefore requires a communication unit with tunable granularity and a stable interface across communication backends.}
\end{tcolorbox}

% \paragraph{A Unified Unit for Intra-Kernel Overlap.}
The combined granularity and backend effects indicate that effective overlap requires a communication unit whose size can match both the rate at which tiles produce data and the efficiency points of different communication backends. At the same time, this unit must provide a stable boundary between the high-level communication schedule and its backend-specific realization, so that schedules need not be rewritten for each backend. We therefore introduce a chunk abstraction that offers both tunable intra-kernel granularity and a unified interface for mapping communication onto diverse backends.

% \paragraph{Summary.}
% The measurements in Figure~\ref{fig:motivation} confirm the limitations of existing kernel-level overlap techniques and highlight the rich but entangled design space that emerges at finer granularity. To exploit this space in practice, we need an automatic method that (1) exposes fine-grained compute--communication overlap without sacrificing SM utilization, and (2) jointly selects communication backends, tile schedules, and chunk sizes. AutoOverlap is designed to meet these requirements.

\section{Overview} \label{sec:overview}

\begin{figure}
    \centering
    \includegraphics[width=0.9\linewidth]{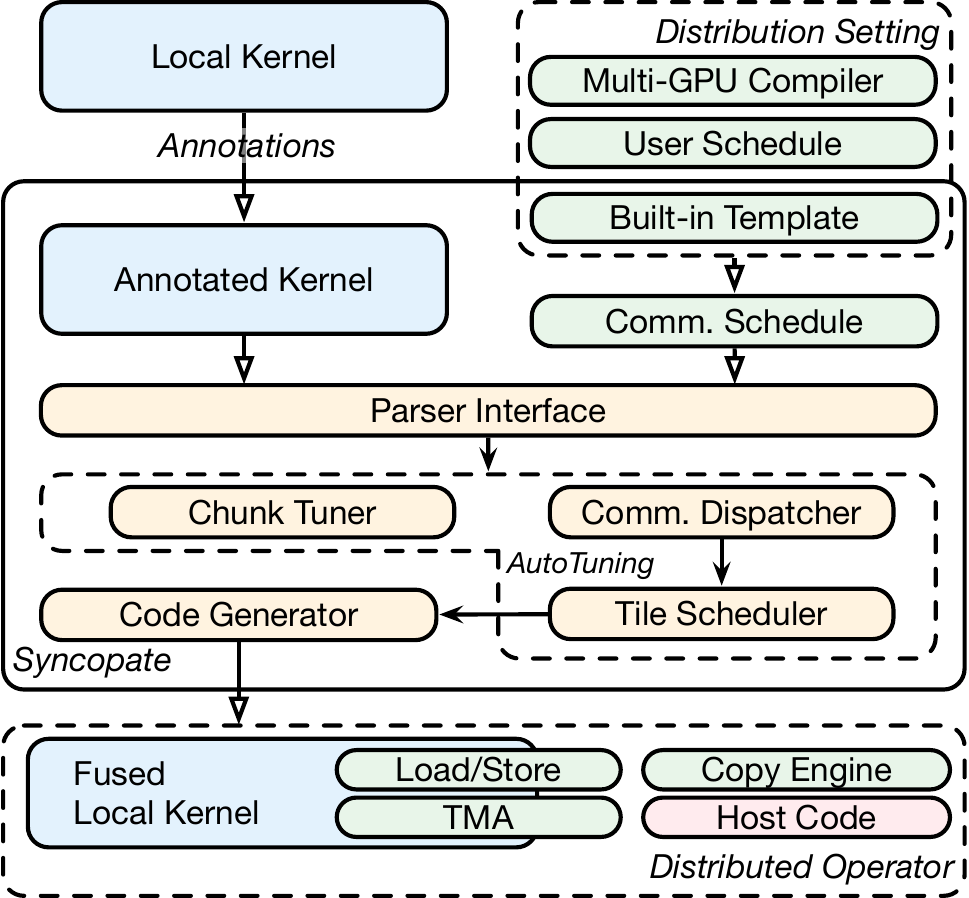}
    \caption{System overview of \thiswork{}.}
    \label{fig:overview}
\end{figure}

\thiswork{} is a compiler and runtime framework that turns locally written Triton kernels into distributed, fine-grained overlapped kernels. Rather than asking programmers to manually fuse communication and computation, \thiswork{} takes an existing local kernel and a high-level distribution specification as input, and automatically synthesizes an intra-kernel schedule that interleaves tile-wise communication with computation according to the available communication backends.

% \begin{CodeListing}[Annotated Local Triton Kernel API]{label=lst:interface}{fontsize=\footnotesize}
% # Original Local Triton Kernel
% @triton.jit
% def kernel_gemm(a_ptr, b_ptr, ...):
%     # Parameters Initialization
%     start_pid = tl.program_id(axis=0)
%     # @ao.axis_count M block=BLOCK_SIZE_M
%     num_pid_m = tl.cdiv(M, BLOCK_SIZE_M)
%     # @ao.tile_id persistent
%     tile_id = start_pid - NUM_SMS
%     ...

%     # Tensor Descriptors
%     a_desc = tl.make_tensor_descriptor(a_ptr, ...)
%     ...

%     # Kernel Main Loop
%     for _ in range(0, k_tiles * tiles_per_SM):
%         tile_id += NUM_SMS

%         # Tile Scheduling
%         # @ao.dispatch begin
%         # @ao.pid_map M=pid_m N=pid_n
%         pid_m, pid_n = get_pid_mn(tile_id, 
%                                   num_pid_m, ...)
%         # @ao.dispatch end

%         offs_am = pid_m * BLOCK_SIZE_M
%         offs_bn = pid_n * BLOCK_SIZE_N
%         offs_k = ki * BLOCK_SIZE_K

%         a = a_desc.load([offs_am, offs_k])
%         b = b_desc.load([offs_bn, offs_k])
%         accumulator = tl.dot(a, b.T, accumulator)
% \end{CodeListing}

\paragraph{Input and User Interface.}
On the compute side, \thiswork{} consumes unmodified local Triton kernels annotated with lightweight scheduling metadata (Listing~\ref{lst:interface}). Programmers write kernels as if they were running on a single device, using standard Triton primitives for indexing, tiling, and tensor descriptors. \rev{Lightweight} \thiswork{} annotations (e.g., axis counts, tile identifiers, and dispatch regions) \rev{are required to} identify the logical tiles and iteration structure but do not change the kernel's semantics. On the distribution side, \thiswork{} uses a communication plan that encodes the desired global data movement pattern and device topology (Listing~\ref{lst:comm_plan}). This plan is expressed using a small API defined by the users or imported directly from higher-level compilers searching parallel schedule.

To end users, \thiswork{} exposes a small set of APIs for (1) registering local kernels with their annotations, (2) constructing or selecting predefined communication plans (such as 1D/2D AllGather or ReduceScatter swizzles), and (3) compiling these into executable distributed kernels. High-level frameworks can wrap these APIs so that model authors only specify tensor partitioning and desired collectives, while \thiswork{} automatically generates the corresponding overlapped kernels. This separation of concerns allows experts to encode distribution and communication strategies once, while ordinary users invoke the resulting distributed operators through familiar, library-style calls.

\begin{figure}[]
    \centering
\begin{minipage}{0.45\textwidth}
\centering
\vspace{-5pt}
\begin{lstlisting}[caption={Annotated Local Triton Kernel API.}, label={lst:interface}]
@triton.jit
def kernel_gemm(a_ptr, b_ptr, ...):
    start_pid = tl.program_id(axis=0)
    # @sy.axis_count M block=BLOCK_SIZE_M
    num_pid_m = tl.cdiv(M, BLOCK_SIZE_M)
    # @sy.tile_id persistent
    tile_id = start_pid - NUM_SMS
    ...
    a_desc = tl.make_tensor_descriptor(a_ptr, ...)
    ...
    for _ in range(0, k_tiles * tiles_per_SM):
        tile_id += NUM_SMS
        # @sy.dispatch begin
        # @sy.pid_map M=pid_m N=pid_n
        pid_m, pid_n = get_pid_mn(tile_id, 
                                  num_pid_m, ...)
        # @sy.dispatch end
        offs_am = pid_m * BLOCK_SIZE_M
        offs_bn = pid_n * BLOCK_SIZE_N
        offs_k = ki * BLOCK_SIZE_K
        a = a_desc.load([offs_am, offs_k])
        b = b_desc.load([offs_bn, offs_k])
        accumulator = tl.dot(a, b.T, accumulator)
\end{lstlisting} 
% \label{lst:interface}
\end{minipage}
\end{figure}

\paragraph{\thiswork{} Architecture and Output.}
Given a local kernel and its communication plan, \thiswork{} lowers them into a unified dependence representation over tiles, shards, and communication operations. The compiler then searches for a fine-grained schedule that maps tiles to devices, assigns communication operations to appropriate backends (e.g., copy engine, TMA, or load/store), and inserts the necessary synchronization to respect both compute and communication dependencies. The output preserves the original numerical semantics while tightly pipelining asynchronous communication and computation around a single fused compute kernel. From the user's perspective, the generated operator can be invoked with the same signature as the original local kernel, plus standard distributed-runtime arguments (e.g., rank, world size, mesh), and integrated into existing training or inference code without further changes.

% \begin{CodeListing}[Communication Schedule Example]{label=lst:comm_plan}{fontsize=\footnotesize}
% def all_gather_1d_swizzle(shape, dtype, axis, rank,...):
%     plan = DevicePlan(dev=rank)
%     plan.tensors_involved[buf] = (torch.Size(shape))

%     local = shard(rank)
%     plan.local_regions.setdefault(buf, []).append(local)

%     for i in range(mesh):
%         peer = (i + rank) % mesh  # 1D swizzle
%         if peer == rank: continue
%         r = shard(peer)
%         plan.add_op(
%             Transfer(
%                 op=TransferOp.PULL,
%                 dst_buf=buf, dst_region=r,
%                 src_buf=buf, src_region=r,
%                 peer=peer, shard_idx=peer,
%                 ...
%             )
%         )
%     return plan
% \end{CodeListing}

\begin{figure}
\centering
\begin{minipage}{0.45\textwidth}
\centering
\vspace{-5pt}
\begin{lstlisting}[caption={Communication Schedule Example.}, label={lst:comm_plan}]
def all_gather_1d_swizzle(shape,dtype,axis,rank,...):
    plan = DevicePlan(dev=rank)
    plan.tensors_involved[buf] = (torch.Size(shape))
    local = shard(rank)
    plan.local_regions.setdefault(buf,[]).append(local)
    for i in range(mesh):
        peer = (i + rank) % mesh  # 1D swizzle
        if peer == rank: continue
        r = shard(peer)
        plan.add_op(Transfer(
                        op=TransferOp.PULL,
                        dst_buf=buf, dst_region=r,
                        src_buf=buf, src_region=r,
                        peer=peer, shard_idx=peer,
                    ...))
    return plan
\end{lstlisting} 
\end{minipage}
% \label{lst:comm_plan}
\end{figure}

\begin{figure*}[]
    \centering
    \includegraphics[width=\linewidth]{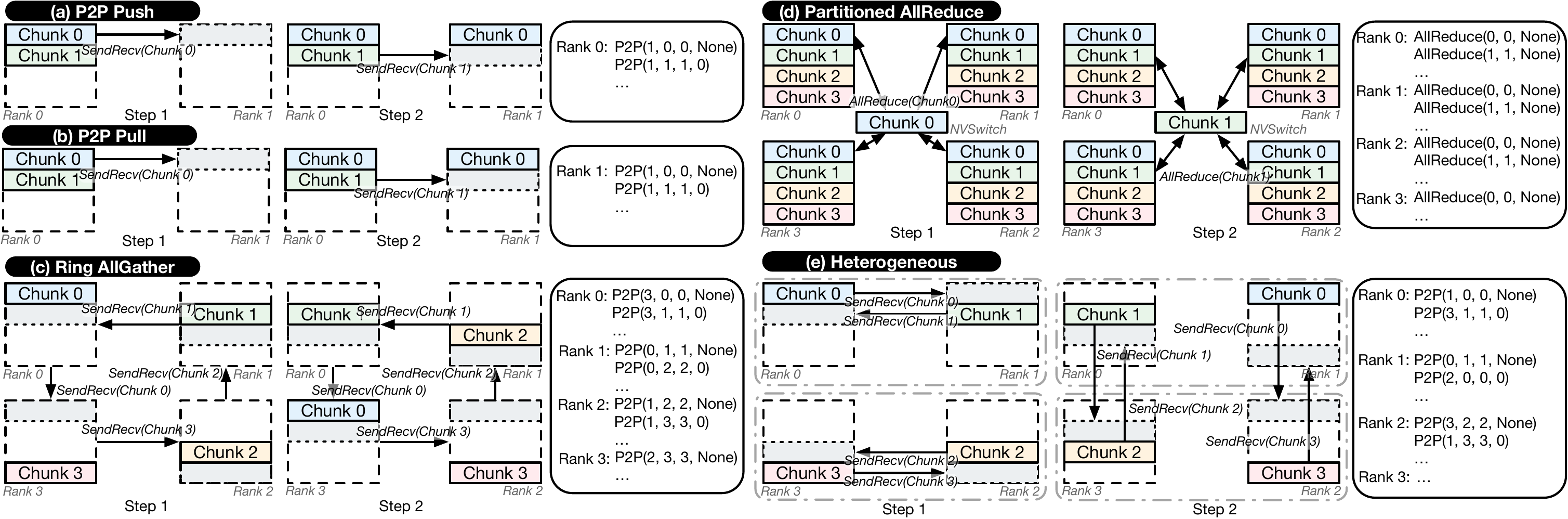}
    \caption{Communication schedule abstraction. (a) and (b) illustrate the same point-to-point exchange expressed as push and pull variants, respectively. (c) shows a ring-based AllGather pattern. (d) represents a partition-based AllReduce schedule. (e) depicts a heterogeneous swizzled AllGather pattern that pipelines communication across multiple hierarchy levels.}
    \label{fig:comm_schedule}
\end{figure*}

\section{\thiswork{}: Fine-Grained Overlap Compiler} \label{sec:method}

\thiswork{} automatically transforms locally written kernels into fine-grained overlapped distributed kernels by aligning the execution of local computation tiles with a global communication schedule. At a high level, we introduce a \emph{chunk-centric} compilation pipeline that treats communication and computation symmetrically: communication is described as transfers of logical chunks, while computation is expressed as tiles that consume and produce these chunks. The compiler then derives dependencies between chunks and tiles, rewrites the tile scheduler to follow the communication order, inserts the necessary synchronization, and finally explores several implementation choices to generate high-performance overlapped code.

\subsection{Communication Schedule Abstraction}

We first define a communication-side abstraction that captures how data moves across devices independently of any particular local kernel implementation. This abstraction is built around the notion of a \emph{chunk}, an intermediate layout between the global logical tensor and the local computation tiles. By operating at this intermediate granularity, the abstraction is expressive enough to describe a wide range of distributed schedules while remaining compatible with both partition-based and loop-based IRs in existing distributed compilers.

\paragraph{Definition.}
A \emph{chunk} is a logical block of data that is communicated as a unit. Each chunk contains one or more tiles, where a tile is the basic unit of computation in the local kernel. Importantly, the chunk size in the communication schedule specifies \emph{logical} transfers; the same logical chunk may later be implemented using different physical communication patterns or backends during lowering.

Conceptually, a chunk can be represented as:
	\texttt{chunk = Chunk(sizes=[...] , layout=..., tensor=...)}.
Based on this abstraction, we define communication operators over chunks. We consider two primary classes of operators:
point-to-point (P2P) transfers and collective communications.

\noindent $\bullet$ \textbf{P2P transfer} is represented as \texttt{P2P(src\_rank, dst\_rank, src\_chunk, dst\_chunk, dependency)}. This operator moves a chunk from a source rank to a destination rank, optionally guarded by a dependency on other chunks or operations. Note that for a pair of P2P operations on the source and destination ranks, we only include the operation on one side. If the P2P operation is defined on the source side, it represents a push operation; otherwise, it represents a pull operation. This will lead to different implementation choices during lowering.

\noindent $\bullet$ \textbf{Collective communication} is represented as \texttt{Collective(collective\_type, src\_chunk, dst\_chunk, ranks, dependency)}. This operator applies a collective operation (e.g., AllGather, ReduceScatter) over a set of ranks on a given chunk with explicit dependency control. When explicitly defined as collective operations, the compiler can leverage the optimized collective implementations provided by the communication backends.

For both operator types, the \texttt{dependency} field encodes any ordering constraints that must be respected between communication operations. Specifically, it is represented as a \texttt{(rank, index)} tuple, indicating that the current operation cannot start until the specified operation on the given rank has completed. This allows us to express complex communication patterns, such as ring exchanges or multi-stage collectives, by chaining dependencies between chunks.

Upon this abstraction, a \emph{communication schedule} is defined as a sequence of chunk-level communication operations with their associated dependencies on each rank as \texttt{schedule := [rank:Int, operations:List[CommOp]]:List}. Since there is no restriction on the operation for each rank, the communication schedule can express heterogeneous communication patterns where different ranks perform different operations on different chunks at different times.

\paragraph{Expressiveness.}
Despite its simplicity, the chunk abstraction is sufficiently expressive to capture a wide range of communication schedules covering all the overlap patterns used in practice (\Fig{\ref{fig:comm_schedule}}). 
To elaborate, (a) and (b) show the same P2P communication between two ranks expressed as push and pull operations, respectively, demonstrating the flexibility of pull/push semantics.
(c) illustrates a ring-based AllGather pattern, which is a common pattern for asynchronous distributed operators\cite{liuringattention}, where each rank sends and receives chunks in a pipelined manner, with dependencies ensuring the correct order of operations. 
(d) depicts a partition-based collective AllReduce pattern, where each rank contributes a chunk to the collective operation and performs the accumulation over the communication fabric. This pattern is often used in partition-based distributed compilers for kernel-level overlap.
Finally, (e) shows a complex heterogeneous swizzled AllGather pattern advancing (c). By utilizing the port abstraction, each rank processes communication at different mesh hierarchy levels, enabling fine-grained pipelining and overlap across multiple dimensions.
With this abstraction, different collective patterns, pipelined P2P exchanges, and hybrid schemes that combine intra-node and inter-node communication can all be written as sequences of chunk-level P2P and collective operators with explicit dependencies. Because chunks are defined in terms of logical tensor regions rather than concrete buffers, the same schedule can be reused across different kernels and tensor shapes, and later specialized by the compiler.

\paragraph{Lowering from Higher-Level Compiler IRs.}
Communication schedules in this abstraction can either be defined manually or automatically derived from existing distributed compiler IRs. 
In the manual case, users construct chunk objects and communication operators directly using our API as shown in Listing~\ref{lst:comm_plan}. \rev{In most cases, manual schedules are parameterized communication algorithms, not fully unrolled per-world-size plans: ring-based algorithms, for example, can be expressed with loops over ranks, chunks, or mesh levels, and \thiswork{} specializes them to the concrete world size, topology, and chunk configuration.}
\thiswork{} also provides pre-defined templates such as 1D/2D AllGather or ReduceScatter swizzles for the common communication patterns.
The user can instantiate these templates with different chunk sizes, mesh topologies, communication axes, and pipeline stages to generate reusable communication schedules.

When integrating with a higher-level distributed compiler, \thiswork{} provides frontends for both partition-based and loop-based IRs.
For partition-based IRs, we analyze the global data partitioning and the implied communication pattern between partitions to infer chunk sizes, participating ranks, and the corresponding P2P or collective operators. For loop-based IRs, we traverse loop nests, identify communication points, and group the communicated regions into chunks according to the chosen granularity. In both cases, the result is a uniform chunk-level schedule that decouples the high-level communication intent from any particular implementation (Listing~\ref{lst:lower-higherir}). Specifically, the collective operators can be directly inserted ("direct") into our communication plan, or they can be further lowered to P2P communication operators using our templates ("template") or using some collective synthesis algorithms ("synth") such as TACOS~\cite{tacos}.

% lowering pesudo code

\begin{figure}[h]
\centering
\begin{minipage}{0.45\textwidth}
\centering
\vspace{-5pt}
\begin{lstlisting}[caption={Lowering from Higher-Level IR.}, label={lst:lower-higherir}]
def emit_steps(steps, mesh, path="template"):
    comm = CommPlan()
    for step in steps:
        if step.is_p2p(): # push/pull/local_copy
            for rank in mesh:
                emit_p2p(comm.plans[rank], step, rank)  
        else:  # collective
            for rank in mesh:
            # Path in ["direct", "synth", "template"]
                emit_collective(path, comm.plans[rank], step, rank)
    return comm

def lower_partition_ir(part_ir, axis_info, mesh, path="template"):
    steps = []
    for tensor in part_ir.tensors:
        layout = part_ir.placement[tensor]
        meta = axis_info[tensor]
        steps.extend(parse_partition_to_steps(tensor, layout, meta))
    return emit_steps(steps, mesh, path)

def lower_loop_ir(loop_ir, mesh, path="template"):
    steps = []
    for node in walk(loop_ir):
        steps.extend(parse_comm_intents(node))
    return emit_steps(steps, mesh, path)
\end{lstlisting} 
% \label{lst:lower-partir}
\end{minipage}
\end{figure}

\subsection{Chunk-based Code Generation}

\begin{figure*}
    \centering
    \includegraphics[width=\linewidth]{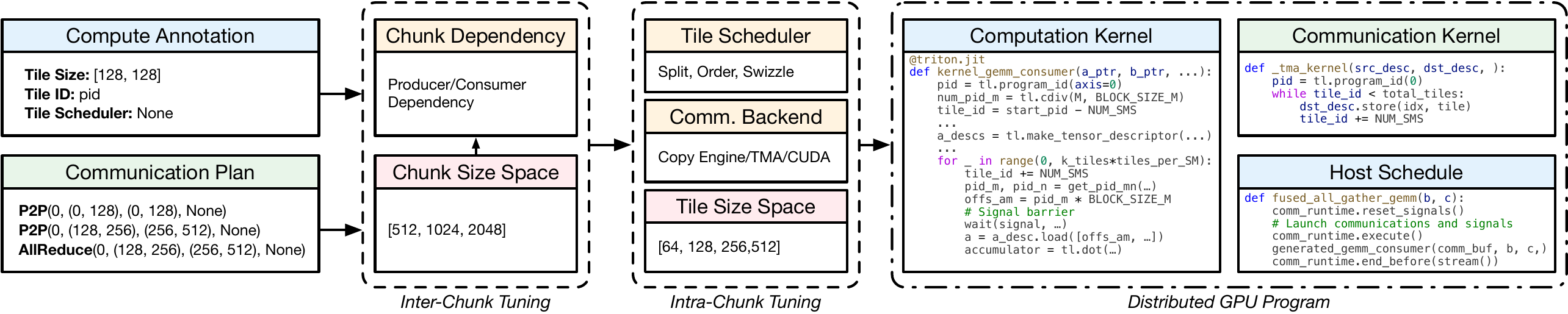}
    \vspace{-2em}
    \caption{Compilation pipeline. In this example, we show communication using specialized SM as an independent kernel synchronized with signals. It can also be a fused kernel depending on the communication backend.}
    \label{fig:workflow}
    \vspace{-5pt}
\end{figure*}

Given a chunk-level communication schedule, the next step is to reorganize local computation so that tiles are executed in an order that aligns with the arrival and consumption of chunks. Intuitively, we want the kernel to compute exactly those tiles whose data has already been communicated, and to defer tiles whose input chunks are still in flight. This \emph{compute chunk scheduling} bridges the gap between the communication abstraction and the tile-level structure of the original local kernel.

\paragraph{Compute Kernel Annotations.}
To make tile-level scheduling explicit, users provide lightweight annotations on the local computation kernel using \thiswork{}'s API. These annotations do not change the numerical semantics of the kernel; instead, they expose its tiling structure and iteration order to the compiler. Although expressed as Python comments, they follow a structured directive format analogous to OpenMP~\cite{openmp} pragmas, allowing the compiler to reliably parse and verify them. Concretely, we require three pieces of information:
% \kz{IIUC, these annotations are expressed as Python comments? I think someone may question how robust this approach could be. It's indeed similar to openmp pragma though}:

\noindent $\bullet$ \textbf{Tile size:} the logical shape of each tile along the relevant dimensions (e.g., GEMM blocks), which allows us to map tiles to chunks.

\noindent $\bullet$ \textbf{Tile index identifier:} a program variable (or tuple of variables) that uniquely identifies the tile being processed in a given iteration.

\noindent $\bullet$ \textbf{Tile scheduler:} the loop or control structure that advances the tile index and determines the order in which tiles are visited.

These annotations can often be derived from existing indexing expressions and loop bounds with minor code changes, as illustrated in the overview section. \rev{
These annotations are compiler metadata. They only expose the tile space, the tile identifier, and the mapping from a tile id to logical tensor-axis coordinates. \thiswork{} uses this information
to relate the local Triton tile scheduler to the chunk-level schedule.
}

\paragraph{Dependency Parsing.}
With both the communication schedule and the annotated compute kernel in hand, \thiswork{} constructs a dependence graph over chunks and tiles \rev{according to the user-specified dependency intent in the plan, including how communication chunks depend on each other and how communication axes map to computation axes.} For each chunk, we track its producer(s) and consumer(s), as well as any explicit ordering constraints encoded in the communication schedule (e.g., pipeline stages). For each tile, we determine which chunks it reads and writes based on its tile index, \rev{tensor layout, and the chunk-to-compute-axis mapping.}

From this graph, the compiler identifies the minimal set of synchronization points needed to respect all data dependencies. Concretely, we insert wait operations in the kernel so that a tile that consumes a given chunk cannot start until the corresponding communication operator has completed. This synchronization can be implemented using different mechanisms depending on the chosen backend, but is always derived from the same chunk-level dependency structure.

\begin{figure}[t]
    \centering
    \includegraphics[width=0.9\linewidth]{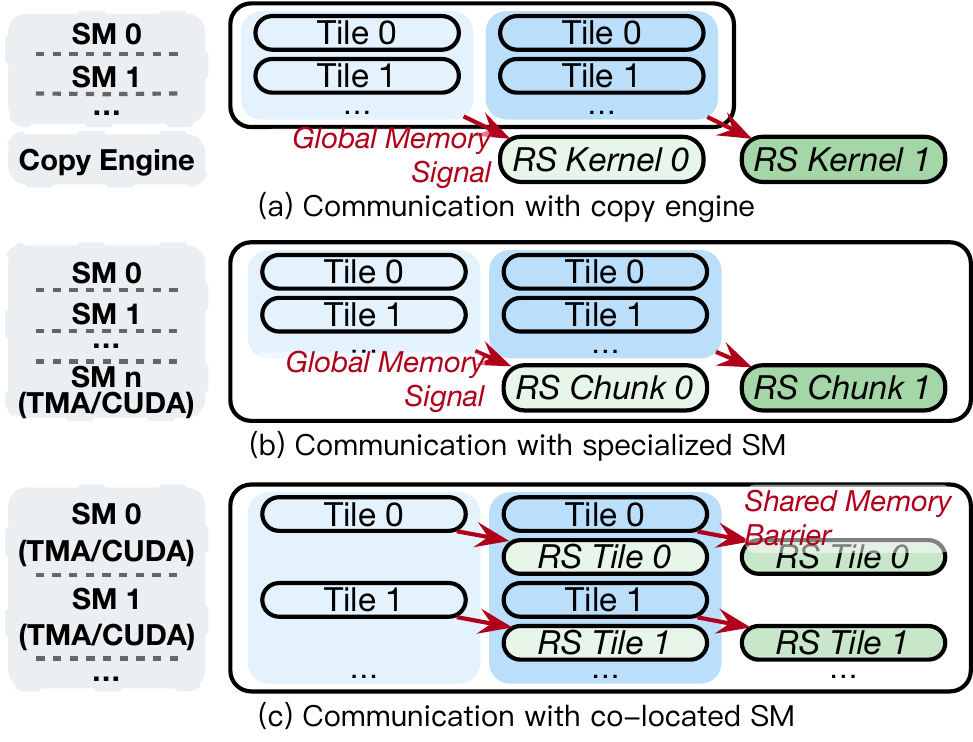}
    \caption{Communication backend selection. (a) Communication issued by the copy engine with global-memory signaling.
(b) Dedicated SMs drive transfer.
(c) Communication is co-located with compute on the same SM.}
    \label{fig:comm_backend}
\end{figure}

\paragraph{Communication Code Generation.}
Once the communication schedule and tile scheduler have been aligned, \thiswork{} lowers the abstract chunk-level plan into concrete communication code. As illustrated in \Fig{\ref{fig:comm_backend}}, the same logical schedule can be realized by several backends that differ in how they move chunks and how they allocate SM resources. Concretely, we support five realizations: (1) using the dedicated copy engine, (2) using TMA on a specialized SM, (3) using TMA on a co-located SM, (4) using operator-instruction load/store on a specialized SM, and (5) using operator-instruction load/store on a co-located SM. In all cases, tiles are produced on one side and consumed by operations on the other side, but the mechanism for signaling readiness and the division of compute versus communication work across SMs differ.

For each operator, \thiswork{} first builds a dependency graph over tiles and communication steps from the chunk schedule, then lowers this graph into backend-specific code that enforces all dependencies by construction. When targeting the copy engine or a specialized SM, the compiler emits global-memory signals and kernel launches so that communication progresses asynchronously relative to the main compute tiles. When targeting co-located SM backends, it instead generates shared-memory barriers and index bookkeeping to coordinate communication and computation within the same SM. Because all five realizations share the same logical schedule but expose different latency/bandwidth and resource trade-offs, they form a search space for the autotuner: \thiswork{} automatically generates all valid implementations, measures their end-to-end performance, and selects the best-performing backend for each operator and hardware configuration.

\begin{figure*}[h]
    \centering
    \includegraphics[width=\linewidth]{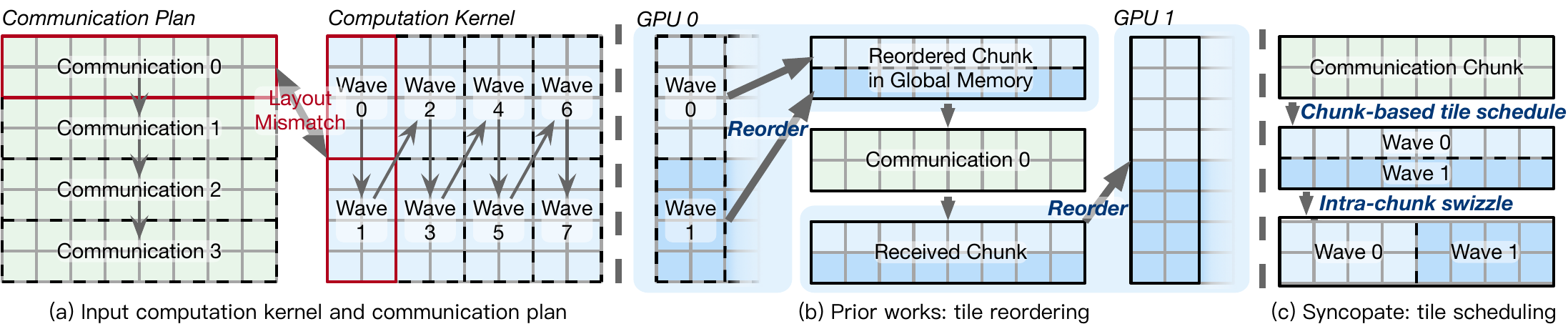}
    \vspace{-2em}
    \caption{Tile scheduler transformation. (a) Computation and communication naturally follow different tile/chunk layouts, creating misalignment. (b) Prior approaches resolve this by inserting explicit data reordering between the two paths. (c) Syncopate instead rewrites the tile schedule to follow chunk order and applies intra-chunk swizzles for locality, aligning compute with communication progress without extra data movement.}
    \label{fig:tile_scheduler}
\end{figure*}

\paragraph{Tile-Scheduler Swizzling.}
As visualized in \Fig{\ref{fig:tile_scheduler}}, the communication plan and the original computation kernel typically induce different layouts over the global tensor: communication groups tiles into chunks based on where data needs to move, while the kernel groups tiles into waves based on its own traversal order. Prior work reconciles this mismatch by explicitly reordering data between communication and computation, paying extra global-memory traffic and synchronization. In contrast, \thiswork{} keeps the communicated chunks in-place and instead \emph{swizzles} the tile scheduler at the intra-kernel level. We reorder the sequence of waves so that each chunk is consumed as soon as it arrives, and apply an intra-chunk swizzle that visits tiles in an order that preserves locality within the chunk. This chunk-based tile schedule aligns compute with communication progress without additional reordering kernels, enabling fine-grained overlap purely through scheduling. \rev{
This transformation has two levels: at the chunk level, \thiswork{} reorders chunks to follow the communication schedule and consume chunks as they become ready; within each chunk, it applies a finer-grained tile scheduler, using the local kernel's original tile traversal by default while allowing alternative axis-based traversals when needed.
}

% Potential benefit: Each CTA waits only once for each arrival chunk; manual writing waits for all tiles.

% persudo code
% for chunks
    % wait
% for tiles
    % wait
\begin{figure*}
    \centering
    \includegraphics[width=\linewidth]{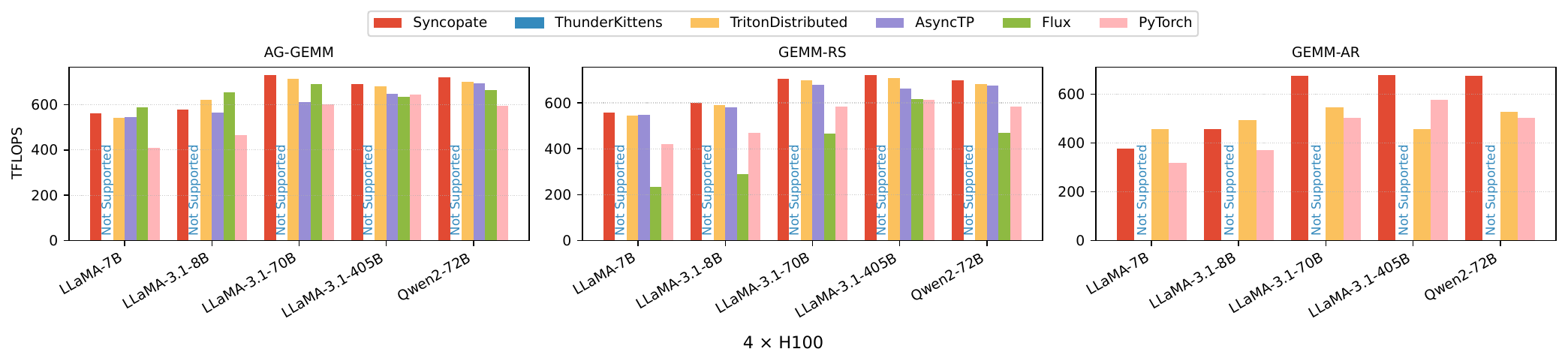}
    \includegraphics[width=\linewidth]{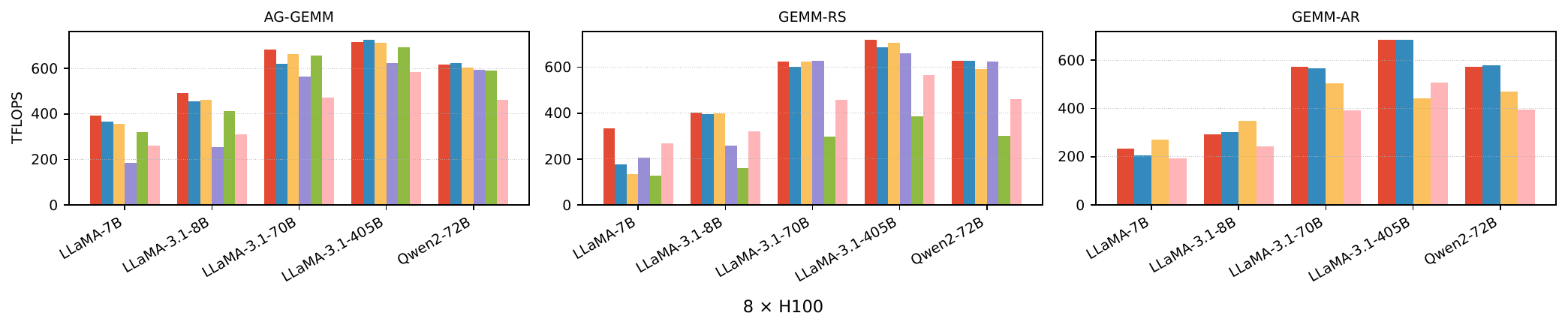}
    \vspace{-3em}
    \caption{Performance comparison of GEMM operators optimized by \thiswork{} with SOTA baselines. ThunderKittens supports only 8 GPUs; 4-GPU is unsupported. When both settings are unsupported, the bar is omitted.}
    \label{fig:operator_results}
\end{figure*}
\subsection{Communication-Centric Auto-Tuning}

The chunk abstraction also provides a natural space for communication-centric auto-tuning. Because chunks sit exactly at the boundary between the global communication schedule and the local tile scheduler, changing chunk-level parameters simultaneously reshapes how data moves across ranks and how computation is ordered within each kernel. Rather than only tuning conventional kernel parameters (e.g., block sizes), \thiswork{} exposes a higher-level search space whose knobs directly control overlap and resource sharing. \rev{Given a local kernel and a chunk-level communication schedule, \thiswork{} uses an enumerate-and-measure tuning pass. The per-candidate compilation overhead is low because \thiswork{} performs lightweight source-to-source rewriting on Triton kernels and relies on the existing Triton JIT backend for code generation.}

At the \emph{inter-chunk} level, we tune the chunk size, shape, and split factor for each logical transfer. Larger chunks tend to achieve higher effective bandwidth on copy engines and TMA but reduce the granularity of overlap, while smaller chunks enable more fine-grained pipelining at the cost of higher per-chunk overhead and more synchronization. Different operators and model sizes favor different trade-off points: communication-heavy A2A-GEMM and GEMM-AR, for example, benefit from intermediate split factors that balance bandwidth and overlap, as confirmed by our sensitivity study in \S\ref{sec:evaluation}. The tuner searches this space under hardware-specific constraints (e.g., minimum efficient transfer size for copy engines and TMA alignment rules) and prunes configurations that would violate these hardware limits.

At the \emph{intra-chunk} level, we tune both the computation tile configuration and how each chunk is realized by a communication backend. Given a fixed logical schedule, the compiler can instantiate each transfer using any of the backends in \Fig{\ref{fig:comm_backend}} (copy engine, intra- or inter-SM TMA, or CUDA load/store on specialized or co-located SMs) and can vary the number of SMs assigned to communication when applicable. Some schedules benefit from using TMA for intra-node tensor movement and load/store-based communication for small, reduction-heavy shards, while others favor copy engines for large bulk transfers with minimal SM involvement. In parallel, the autotuner explores different tile sizes and intra-tile orders that better align compute waves with the chosen chunk layout, improving locality and avoiding long communication tails.

Crucially, all of these decisions operate on top of the same chunk-level dependence graph. Changing the backend, SM allocation, or tile order never requires re-deriving the global communication plan; instead, \thiswork{} reuses the existing schedule and regenerates backend-specific code that enforces the same dependencies. This separation of logical schedule from physical realization is what makes the search space both rich and manageable: as our ablation results show, reasonable but suboptimal settings can easily leave more than a factor of two in performance, while the tuned configuration found by our communication-centric autotuner consistently coincides with the most balanced point between computation, communication, and hardware utilization.

\section{Evaluation} \label{sec:evaluation}

\subsection{Experimental Setup}

\paragraph{Testbed.}
We evaluate \thiswork{} on a server with 8 NVIDIA H100 GPUs connected via NVLink with an aggregate bandwidth of 900 GB/s, which is a representative single-node setting used in prior multi-GPU kernel studies~\cite{chang2024flux, zheng2025tilelink, parallelkitten, mercury}. Unless otherwise stated, all measurements are taken on a single node using all 8 GPUs; in later experiments, we vary the number of active devices to study scalability and portability. \thiswork{} is implemented with CUDA~v12.9, NVSHMEM~v3.3.9, and PyTorch~v2.7, and we run all baselines on the same software stack to ensure a fair comparison.

\paragraph{Workloads.}
We use \thiswork{} to optimize representative multi-GPU operators that dominate the cost of modern LLM workloads: general matrix multiplication (GEMM) and attention~\cite{vaswani2017attention, dao2022flashattention, shah2024flashattention, dao2023flashattention}. For GEMM, we benchmark three distributed variants: AllGather--GEMM (AG-GEMM) and GEMM--ReduceScatter (GEMM-RS), and GEMM--AllReduce (GEMM-AR), which appear in tensor-parallel~\cite{shoeybi2019megatron} or sequence parallel~\cite{korthikanti2023reducing} feed-forward network (FFN) layers. For attention, we evaluate both head-parallel (HP)~\cite{jacobs2023deepspeed} and sequence-parallel (SP) schedules, including the overlapped RingAttention (Ring-Attn)~\cite{liuringattention} variant.

Operator shapes are derived from the FFN layers and attention layers of open-source Llama-3\cite{grattafiori2024llama} and Qwen~\cite{qwen2025qwen25technicalreport} models, covering a range of hidden dimensions, head counts, and parallelism configurations that are typical of large-scale LLM deployments. For attention, we sweep over multiple sequence lengths to reflect common short- and long-context use cases under different distribution strategies. Overall, this workload suite exercises \thiswork{} across both regular GEMM-heavy and more irregular attention patterns.

\begin{figure*}
    \centering
    \includegraphics[width=\linewidth]{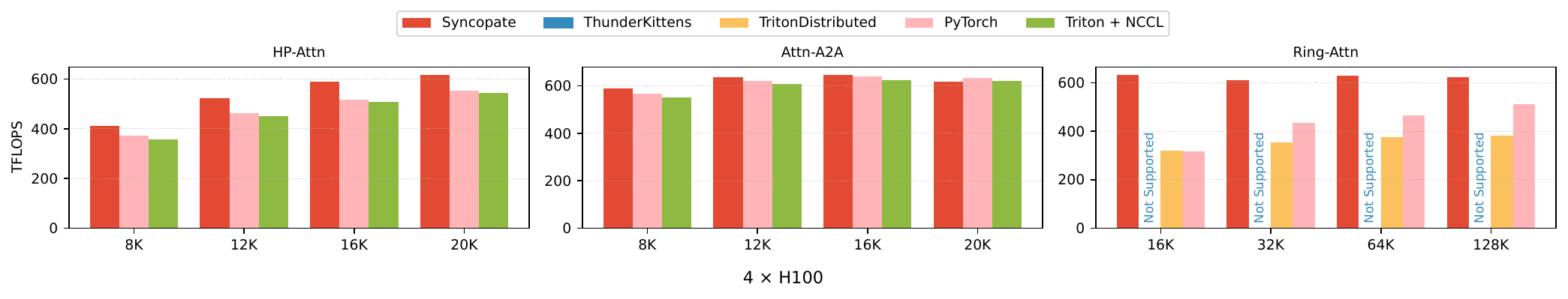}
    \includegraphics[width=\linewidth]{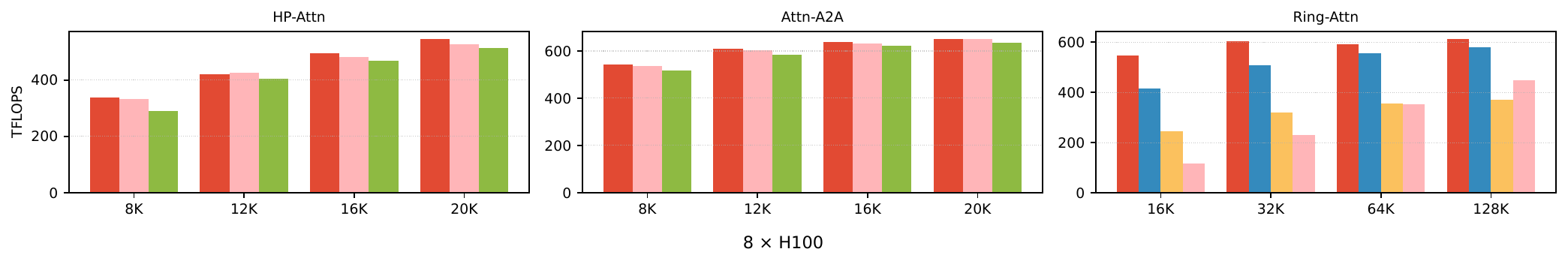}
    \vspace{-2.5em}
    \caption{Performance comparison of operators optimized by \thiswork{} with SOTA baselines.}
    \label{fig:attention_results}
    \vspace{-0.4cm}
\end{figure*}

\paragraph{Baselines.}
To assess the effectiveness of operators generated and tuned by \thiswork{}, we compare against both state-of-the-art manually engineered kernels and fully automatic compiler-based approaches. As manual baselines, we include built-in operators from fine-grained overlap DSLs such as ThunderKittens~\cite{thunderkitten, parallelkitten} and TritonDistributed~\cite{zheng2025triton, zheng2025tilelink}, as well as highly optimized implementations including AsyncTP\cite{OverlapCommunicationwithDependentComputationviaDecomposition}, Flux\cite{chang2024flux}, and Triton kernels paired with NCCL~\cite{nccl} collectives.

To isolate the benefit of \thiswork{}'s automatic fine-grained overlap, we further compare against automatic operators produced by existing distributed compiler frameworks, including Domino\cite{wang2024dominoeliminatingcommunicationllm}, Alpa\cite{zheng2022alpa}, and Mercury\cite{mercury}. For these comparisons, we transform the communication schedules found by each compiler into our chunk-level representation and reuse the same high-level plans in \thiswork{}, so that any performance difference reflects our intra-kernel overlap and backend-selection mechanisms rather than differences in global parallelization strategy.

\rev{
\paragraph{Overlap Granularity.}
AsyncTP, Flux, TritonDistributed, and ThunderKittens perform fine-grained intra-kernel overlap by explicitly coordinating communication and computation inside specialized kernels or DSL programs.
Domino, Mercury, and the PyTorch RingAttention baseline use inter-kernel overlap, while the remaining baselines execute communication through separate library collectives without overlap.
}

\rev{
\paragraph{Base Triton Kernels.}
Our GEMM local kernels are off-the-shelf Triton GEMM kernels with only \thiswork{} annotations added.
For attention, we start from standard Triton attention kernels and make minor modifications to support split-KV execution, which is not exposed by the original kernels.
In both cases, the kernels remain local compute kernels; \thiswork{} generates the communication logic, synchronization, backend selection, and tile-schedule transformations.
}

% \begin{itemize}
%     \item Main results: GEMM, Attention (Baselines?)
%     \item (Backend results: AMD?)
%     \item (Schedule Sensitivity)
%     \item Communication Backend Abalation
%     \item Integration Experiment
%     \item Design Space (Tile, Chunk Size tuning)
% \end{itemize}

\subsection{Performance Benchmark}

\paragraph{Operator Results.}
Across all evaluated settings, \thiswork{} generates automatically optimized operators \rev{that are competitive with} carefully hand-engineered baselines, while showing even larger gains over fully automatic distributed compilers. As summarized in \Fig{\ref{fig:operator_results}} and \Fig{\ref{fig:attention_results}}, \thiswork{} sustains high TFLOPS on both GEMM and attention operators under multiple communication patterns (AG-GEMM, GEMM-RS, GEMM-AR, HP/SP attention, and Ring-Attn) and model configurations derived from Llama-3 and Qwen.

In \Fig{\ref{fig:operator_results}}, \thiswork{} is at or near the best-performing curve in almost every GEMM configuration. On common, heavily optimized AG-GEMM and GEMM-RS cases where manual kernels such as ThunderKittens, TritonDistributed, AsyncTP, and Flux already sit close to the hardware limits, \thiswork{} essentially matches their peak throughput on both 4- and 8-GPU settings, achieving on average 99.8\% of the best baseline on 4 GPUs and 104\% on 8 GPUs. For GEMM-AR, \thiswork{} is marginally below TritonDistributed on 7B/8B shapes, yet it scales more effectively and becomes the top-performing kernel on larger model configurations. These results indicate that our generic compilation pipeline can recover the same highly tuned overlap patterns that experts design by hand. \rev{
The small advantage over the strongest baseline in some 8-GPU cases comes from
the chunk-level abstraction. By grouping
multiple tiles under a communication chunk, \thiswork{} reduces the number of
fine-grained readiness checks and synchronization points while still preserving
overlap.
}

In \Fig{\ref{fig:attention_results}}, we observe a similar trend for attention operators. Under standard HP attention with moderate sequence lengths, \thiswork{} closely tracks the best manual implementations, confirming that the chunk-based abstraction does not sacrifice performance even for workloads that already benefit from hand-optimized kernels. As the problem becomes harder, moving to Ring-Attn, longer sequences, and 8-GPU runs, the gap widens in favor of \thiswork{}. Our compiler maintains high TFLOPS while baseline kernels degrade more rapidly, since it can reshape chunks, rebalance compute and communication, and choose different backends as the sequence length and parallelism strategy change. Notably, on the most communication-intensive Ring-Attn settings, \thiswork{} delivers the best performance despite not being tailored specifically for this operator.

\begin{figure}
    \centering
    \includegraphics[width=0.49\linewidth]{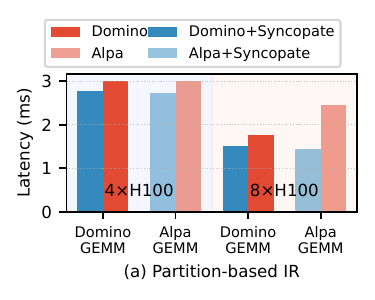}~
    \includegraphics[width=0.49\linewidth]{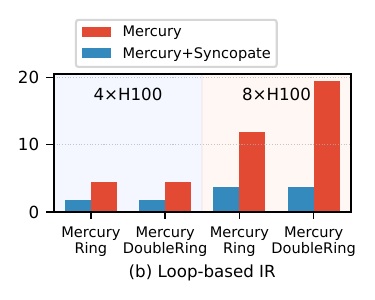}
    \caption{Evaluation of lowering partition-based and loop-based IRs generated by higher-level distributed compilers into \thiswork{}. }
    \label{fig:ir_example}
\end{figure}

\paragraph{Integration Results.}
We further evaluate how \thiswork{} composes with existing automatic distributed compilers that search for the communication schedule among devices, using their communication plans as our inputs. As summarized in \Fig{\ref{fig:ir_example}}, for each of Domino, Alpa, and Mercury, we keep the original parallelization strategy and the searched communication schedule fixed, convert that schedule into our chunk-level representation, and let \thiswork{} generate the fine-grained overlapped kernels. Across both GEMM and attention workloads on 4- and 8-H100 configurations, integrating \thiswork{} consistently reduces end-to-end operator latency compared to the native implementations shipped with these systems, showing that chunk-based intra-kernel overlap exposes an additional optimization dimension on top of their global parallelization decisions.

This experiment also illustrates that \thiswork{} can be cleanly integrated with both partition-based and loop-based compiler stacks. Domino and Alpa operate on partitioned IRs and expose their communication plans as sequences of collectives between logical tensor partitions, while Mercury is built around a loop-centric IR for ring~\cite{liuringattention} and double-ring attention~\cite{gu2024loongtrain}. Because our interface only requires a well-defined, implementation-agnostic communication schedule and light-weight annotations on the local kernels, these systems can plug into \thiswork{} with minimal source modifications. This compatibility allows practitioners to reuse mature distributed compilers for global partitioning, while relying on \thiswork{} to automatically realize high-performance, fine-grained overlap within each generated operator.

% \begin{figure}
%     \centering
%     \includegraphics[width=\linewidth]{figure/backend_ablation.pdf}
%     \caption{Ablation study on communication backend selection.}
%     \label{fig:backend_ablation}
% \end{figure}

% \begin{figure}
%     \centering
%     \includegraphics[width=\linewidth]{figure/chunk_size_sensitivity.pdf}
%     \caption{Sensitivity analysis of chunk size on operator performance.}
%     \label{fig:chunk_size_sensitivity}
% \end{figure}

% \begin{figure}
%     \centering
%     \includegraphics[width=\linewidth]{figure/tile_size_sensitivity.pdf}
%     \caption{Sensitivity analysis of tile size on operator performance.}
%     \label{fig:tile_size_sensitivity}
% \end{figure}

% \begin{figure}
%     \centering
%     \includegraphics[width=\linewidth]{figure/sm_sensitivity.pdf}
%     \caption{Sensitivity analysis of the number of active SMs on operator performance.}
%     \label{fig:sm_sensitivity}
% \end{figure}

\subsection{Ablation and Sensitivity Studies}

\begin{figure*}
    \includegraphics[width=0.49\linewidth]{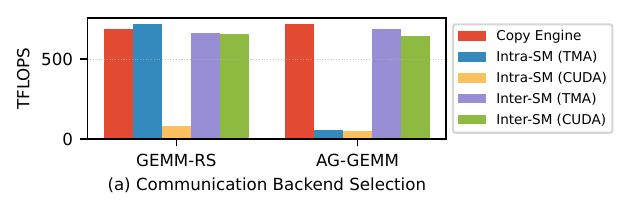}~
    \includegraphics[width=0.49\linewidth]{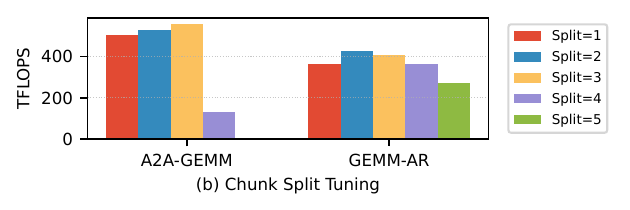}
    \includegraphics[width=0.49\linewidth]{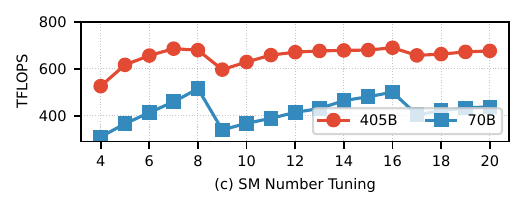}~
    \includegraphics[width=0.49\linewidth]{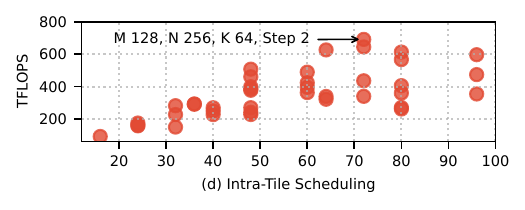}
    \caption{Ablation and sensitivity studies of \thiswork{}'s auto-tuning design space.}
    \label{fig:ablation_sensitivity}
\end{figure*}

We next evaluate the effect of \thiswork{}'s chunk-based code generation and communication-centric auto-tuning components through a series of ablation and sensitivity analyses in \Fig{\ref{fig:ablation_sensitivity}}. Each subplot corresponds to one of the key design choices introduced in \S\ref{sec:method}: communication backend selection and SM allocation, chunk size (split factor), and intra-tile scheduling.

\paragraph{Communication Backend and SM Tuning.}
\Fig{\ref{fig:ablation_sensitivity}}(a) and (c) study how different realizations of the same logical communication schedule perform under the backends described in \S\,5.2, and how tuning the number of active SMs further refines this choice. In (a), for GEMM-RS and AG-GEMM, copy-engine and intra-SM TMA backends achieve the highest TFLOPS, while purely CUDA load/store realizations saturate at much lower throughput. The gap between the best and worst backend for the same logical schedule is comparable to the gap between our final implementation and several baselines in \Fig{\ref{fig:operator_results}}, indicating that backend selection alone can determine whether overlap is effective. This confirms that the ability to instantiate the same chunk schedule with different backends is crucial: no single mechanism dominates across operators, and picking a suboptimal backend can leave more than half of the available performance on the table. In (c), for a fixed backend, varying the number of SMs devoted to communication reveals a clear sweet spot where computation and communication are balanced. For both 405B and 70B GEMMs, allocating too few SMs underutilizes the link bandwidth, whereas allocating too many starves the main kernel. The optimal SM count also shifts with model size, which matches the design of our backend code generation: the autotuner treats SM allocation as a first-class knob and automatically selects a near-optimal point for each operator/hardware pair instead of relying on a single hard-coded ratio.

\paragraph{Chunk Size Tuning.}
\Fig{\ref{fig:ablation_sensitivity}}(b) varies the number of chunks (split factor) used to realize the same high-level schedule for A2A-GEMM and GEMM-AR. Smaller split factors correspond to larger chunks with higher single-transfer efficiency but fewer overlap opportunities, while larger split factors increase overlap at the cost of per-chunk overhead. The curves exhibit a clear non-monotonic trend, with performance peaking at an intermediate split (e.g., 2--3 splits, about 128MB for GEMM-AR) and degrading when chunks become either too coarse or too fine. Notably, naive choices that are convenient to implement by hand (such as a single large chunk or splitting once per rank) sit far from the optimum. This behavior directly reflects the trade-off discussed in our chunk-based code generation: practitioners cannot reliably pick a single ``good'' chunk size by hand, whereas \thiswork{} searches this space automatically and selects the configuration that best matches the operator's compute/communication balance.

\paragraph{Intra-Tile Scheduling.}
Finally, \Fig{\ref{fig:ablation_sensitivity}}(d) explores different intra-tile scheduling strategies for a representative GEMM configuration, varying the order in which tiles within each chunk are visited. Each point corresponds to a valid schedule with tile size on M, N, K dimensions and the pipeline stages that preserve program semantics but change locality and load balance. To represent different search candidates, we calculate the consumed shared memory size and plot the performance of these valid schedules. The wide spread in TFLOPS shows that tile order alone can introduce more than a 2\,\texttimes{} performance difference, reinforcing the importance of the tile-scheduler transformation in \S\,5.2. High-performing schedules cluster around orders that align tile waves with the communication chunk order introduced in our chunk-based code generation, whereas poorly performing ones repeatedly revisit tiles in a way that destroys locality or creates long tails of unfinished work. By generating and evaluating these schedules automatically, \thiswork{} converges on tile orders that co-optimize cache reuse and SM utilization, without requiring users to reason manually about low-level tiling and swizzling policies.

Taken together, these studies demonstrate that \thiswork{}'s chunk abstraction is not only expressive but also forms a practical search space: the same logical schedule can be realized via multiple backends, chunk sizes, SM allocations, and tile orders, and the differences between reasonable but non-optimal choices and the tuned configuration are often comparable to, or larger than, the gaps between systems in our main benchmarks. The auto-tuning framework in \S\,5.3 systematically explores this space using the code generation mechanisms of \S\,5.2, turning what would otherwise be a brittle, hand-tuned process into a robust, compiler-driven optimization.

\section{Discussion and Future Work}

\paragraph{Extending to other DSLs.} \rev{
\thiswork{} is implemented as a Triton source-to-source compiler, but the core
abstraction only requires a tile-centric frontend that exposes tile shapes, tile
identifiers, tile traversal order, and tile memory regions. Thus, DSLs such as
CuTeDSL~\cite{nvidia_cute_dsl} or cuTile~\cite{nvidia_cutile} could be supported with an additional frontend/lowering layer,
as long as kernels are written in a tiled style. The chunk schedules,
dependency construction, and backend-selection logic would remain reusable.}

\paragraph{Scope of compute-memory overlap.} \rev{\thiswork{} targets fine-grained overlap between tiled computation and explicit inter-GPU communication.
Overlapping independent compute-bound and memory-bound kernels, as in NanoFlow-style optimizations~\cite{zhu2025nanoflowoptimallargelanguage}, is largely orthogonal and can often be handled by stream-level scheduling.
Such optimizations could be modeled by \thiswork{} only when the stages must be coordinated as chunked producer-consumer relationships with tile-level dependencies.}

\paragraph{Dynamically shaped workloads.} \rev{\thiswork{} currently specializes kernels to a fixed operator shape or a small set of common shape buckets. For variable sequence lengths, the chunk schedule can usually be re-instantiated with the new shape parameters; when chunk ranges are stored as device-side metadata, a lightweight update kernel can refresh these values before launching the generated kernel. A full dynamic-shape runtime that automatically manages shape buckets and metadata updates is left to future work.}

\paragraph{Multi-node settings.} \rev{
\thiswork{}'s chunk abstraction naturally extends to multi-node settings, since larger chunks are also a suitable unit for bandwidth-efficient NIC transfers. The main extension is not to generalize tiles or chunks, but to enrich the communication plan with a channel dimension that distinguishes intra-node and inter-node paths. Different channels can progress concurrently, allowing NVLink transfers and NIC transfers to be scheduled and overlapped independently. The compiler can then assign chunks to channels according to the topology and use different backend realizations for each channel. Supporting this setting requires additional backend and runtime support for inter-node communication, which we leave to future work.
}

\paragraph{Why a compiler abstraction?} \rev{
\thiswork{} is not meant to replace expert-written kernels for a fixed operator and hardware target. Its goal is to avoid re-encoding the same fine-grained overlap logic, signal/wait protocol, and backend-specific communication code for each schedule or parallel configuration. The chunk abstraction lets users or higher-level compilers specify the logical communication policy once, while \thiswork{} lowers it to concrete dependencies, synchronization, backend choices, and tile schedules.
}
\section{Conclusion} \label{sec:conclusion}

Syncopate bridges the abstraction gap between communication planning and fine-grained overlapping by introducing a chunk-based interface that unifies communication plans with tiled GPU computation. By automatically aligning tile schedules with communication progress and selecting among heterogeneous backends, Syncopate turns intra-kernel overlap into a general compiler capability rather than a hand-engineered optimization. The framework integrates cleanly with existing distributed compilers, complementing their global parallelization strategies with backend-aware, fine-grained scheduling. This decoupled design provides a foundation for systems that co-schedule computation and communication and accommodate diverse communication backends.

\bibliographystyle{plain}
\bibliography{reference}

@article{shoeybi2019megatron,
  title={Megatron-lm: Training multi-billion parameter language models using model parallelism},
  author={Shoeybi, Mohammad and Patwary, Mostofa and Puri, Raul and LeGresley, Patrick and Casper, Jared and Catanzaro, Bryan},
  journal={arXiv preprint arXiv:1909.08053},
  year={2019}
}

@article{korthikanti2023reducing,
  title={Reducing activation recomputation in large transformer models},
  author={Korthikanti, Vijay Anand and Casper, Jared and Lym, Sangkug and McAfee, Lawrence and Andersch, Michael and Shoeybi, Mohammad and Catanzaro, Bryan},
  journal={Proceedings of Machine Learning and Systems},
  volume={5},
  pages={341--353},
  year={2023}
}

@article{vaswani2017attention,
  title={Attention is all you need},
  author={Vaswani, Ashish and Shazeer, Noam and Parmar, Niki and Uszkoreit, Jakob and Jones, Llion and Gomez, Aidan N and Kaiser, {\L}ukasz and Polosukhin, Illia},
  journal={Advances in neural information processing systems},
  volume={30},
  year={2017}
}

@inproceedings{zheng2022alpa,
  title={Alpa: Automating inter-and $\{$Intra-Operator$\}$ parallelism for distributed deep learning},
  author={Zheng, Lianmin and Li, Zhuohan and Zhang, Hao and Zhuang, Yonghao and Chen, Zhifeng and Huang, Yanping and Wang, Yida and Xu, Yuanzhong and Zhuo, Danyang and Xing, Eric P and others},
  booktitle={16th USENIX Symposium on Operating Systems Design and Implementation (OSDI 22)},
  pages={559--578},
  year={2022}
}

@inproceedings{chen2024centauri,
  title={Centauri: Enabling efficient scheduling for communication-computation overlap in large model training via communication partitioning},
  author={Chen, Chang and Li, Xiuhong and Zhu, Qianchao and Duan, Jiangfei and Sun, Peng and Zhang, Xingcheng and Yang, Chao},
  booktitle={Proceedings of the 29th ACM International Conference on Architectural Support for Programming Languages and Operating Systems, Volume 3},
  pages={178--191},
  year={2024}
}

@inproceedings{liuringattention,
  title={RingAttention with Blockwise Transformers for Near-Infinite Context},
  author={Liu, Hao and Zaharia, Matei and Abbeel, Pieter},
  booktitle={The Twelfth International Conference on Learning Representations}
}

@inproceedings{chen2018tvm,
  title={$\{$TVM$\}$: An automated $\{$End-to-End$\}$ optimizing compiler for deep learning},
  author={Chen, Tianqi and Moreau, Thierry and Jiang, Ziheng and Zheng, Lianmin and Yan, Eddie and Shen, Haichen and Cowan, Meghan and Wang, Leyuan and Hu, Yuwei and Ceze, Luis and others},
  booktitle={13th USENIX Symposium on Operating Systems Design and Implementation (OSDI 18)},
  pages={578--594},
  year={2018}
}

@article{gu2024loongtrain,
  title={Loongtrain: Efficient training of long-sequence llms with head-context parallelism},
  author={Gu, Diandian and Sun, Peng and Hu, Qinghao and Huang, Ting and Chen, Xun and Xiong, Yingtong and Wang, Guoteng and Chen, Qiaoling and Zhao, Shangchun and Fang, Jiarui and others},
  journal={arXiv preprint arXiv:2406.18485},
  year={2024}
}

@article{chang2024flux,
  title={FLUX: fast software-based communication overlap on gpus through kernel fusion},
  author={Chang, Li-Wen and Bao, Wenlei and Hou, Qi and Jiang, Chengquan and Zheng, Ningxin and Zhong, Yinmin and Zhang, Xuanrun and Song, Zuquan and Yao, Chengji and Jiang, Ziheng and others},
  journal={arXiv preprint arXiv:2406.06858},
  year={2024}
}

@inproceedings{jangda2022breaking,
  title={Breaking the computation and communication abstraction barrier in distributed machine learning workloads},
  author={Jangda, Abhinav and Huang, Jun and Liu, Guodong and Sabet, Amir Hossein Nodehi and Maleki, Saeed and Miao, Youshan and Musuvathi, Madanlal and Mytkowicz, Todd and Saarikivi, Olli},
  booktitle={Proceedings of the 27th ACM International Conference on Architectural Support for Programming Languages and Operating Systems},
  pages={402--416},
  year={2022}
}

@article{santhanam2021distir,
  title={Distir: An intermediate representation and simulator for efficient neural network distribution},
  author={Santhanam, Keshav and Krishna, Siddharth and Tomioka, Ryota and Harris, Tim and Zaharia, Matei},
  journal={arXiv preprint arXiv:2111.05426},
  year={2021}
}

@inproceedings{alabed2025partir,
  title={PartIR: Composing SPMD Partitioning Strategies for Machine Learning},
  author={Alabed, Sami and Belov, Daniel and Chrzaszcz, Bart and Franco, Juliana and Grewe, Dominik and Maclaurin, Dougal and Molloy, James and Natan, Tom and Norman, Tamara and Pan, Xiaoyue and others},
  booktitle={Proceedings of the 30th ACM International Conference on Architectural Support for Programming Languages and Operating Systems, Volume 1},
  pages={794--810},
  year={2025}
}

@article{jacobs2023deepspeed,
  title={Deepspeed ulysses: System optimizations for enabling training of extreme long sequence transformer models},
  author={Jacobs, Sam Ade and Tanaka, Masahiro and Zhang, Chengming and Zhang, Minjia and Song, Shuaiwen Leon and Rajbhandari, Samyam and He, Yuxiong},
  journal={arXiv preprint arXiv:2309.14509},
  year={2023}
}

@inproceedings{triton,
author = {Tillet, Philippe and Kung, H. T. and Cox, David},
title = {Triton: an intermediate language and compiler for tiled neural network computations},
year = {2019},
isbn = {9781450367196},
publisher = {Association for Computing Machinery},
address = {New York, NY, USA},
url = {https://doi.org/10.1145/3315508.3329973},
doi = {10.1145/3315508.3329973},
booktitle = {Proceedings of the 3rd ACM SIGPLAN International Workshop on Machine Learning and Programming Languages},
pages = {10–19},
numpages = {10},
location = {Phoenix, AZ, USA},
series = {MAPL 2019}
}

@inproceedings{pytorch2,
author = {Ansel, Jason and Yang, Edward and He, Horace and Gimelshein, Natalia and Jain, Animesh and Voznesensky, Michael and Bao, Bin and Bell, Peter and Berard, David and Burovski, Evgeni and Chauhan, Geeta and Chourdia, Anjali and Constable, Will and Desmaison, Alban and DeVito, Zachary and Ellison, Elias and Feng, Will and Gong, Jiong and Gschwind, Michael and Hirsh, Brian and Huang, Sherlock and Kalambarkar, Kshiteej and Kirsch, Laurent and Lazos, Michael and Lezcano, Mario and Liang, Yanbo and Liang, Jason and Lu, Yinghai and Luk, C. K. and Maher, Bert and Pan, Yunjie and Puhrsch, Christian and Reso, Matthias and Saroufim, Mark and Siraichi, Marcos Yukio and Suk, Helen and Zhang, Shunting and Suo, Michael and Tillet, Phil and Zhao, Xu and Wang, Eikan and Zhou, Keren and Zou, Richard and Wang, Xiaodong and Mathews, Ajit and Wen, William and Chanan, Gregory and Wu, Peng and Chintala, Soumith},
title = {PyTorch 2: Faster Machine Learning Through Dynamic Python Bytecode Transformation and Graph Compilation},
year = {2024},
isbn = {9798400703850},
publisher = {Association for Computing Machinery},
address = {New York, NY, USA},
url = {https://doi.org/10.1145/3620665.3640366},
doi = {10.1145/3620665.3640366},
booktitle = {Proceedings of the 29th ACM International Conference on Architectural Support for Programming Languages and Operating Systems, Volume 2},
pages = {929–947},
numpages = {19},
location = {La Jolla, CA, USA},
series = {ASPLOS '24}
}

@inproceedings{feng2023tensorir,
  title={Tensorir: An abstraction for automatic tensorized program optimization},
  author={Feng, Siyuan and Hou, Bohan and Jin, Hongyi and Lin, Wuwei and Shao, Junru and Lai, Ruihang and Ye, Zihao and Zheng, Lianmin and Yu, Cody Hao and Yu, Yong and others},
  booktitle={Proceedings of the 28th ACM International Conference on Architectural Support for Programming Languages and Operating Systems, Volume 2},
  pages={804--817},
  year={2023}
}

@manual{nccl,
  title        = {NVIDIA Collective Communications Library (NCCL)},
  author       = {{NVIDIA Corporation}},
  year         = {2025},
  note         = {Version 2.26.2},
  url          = {https://developer.nvidia.com/nccl}
}

@article{dao2022flashattention,
  title={Flashattention: Fast and memory-efficient exact attention with io-awareness},
  author={Dao, Tri and Fu, Dan and Ermon, Stefano and Rudra, Atri and R{\'e}, Christopher},
  journal={Advances in neural information processing systems},
  volume={35},
  pages={16344--16359},
  year={2022}
}

@article{dao2023flashattention,
  title={Flashattention-2: Faster attention with better parallelism and work partitioning},
  author={Dao, Tri},
  journal={arXiv preprint arXiv:2307.08691},
  year={2023}
}

@article{shah2024flashattention,
  title={Flashattention-3: Fast and accurate attention with asynchrony and low-precision},
  author={Shah, Jay and Bikshandi, Ganesh and Zhang, Ying and Thakkar, Vijay and Ramani, Pradeep and Dao, Tri},
  journal={Advances in Neural Information Processing Systems},
  volume={37},
  pages={68658--68685},
  year={2024}
}

@techreport{nvidia_nvlink_whitepaper,
  title        = {NVIDIA NVLink High-Speed Interconnect: Application Performance},
  author       = {{NVIDIA Corporation}},
  year         = {2015},
  institution  = {NVIDIA Corporation},
  url          = {https://info.nvidianews.com/rs/nvidia/images/NVIDIA%20NVLink%20High-Speed%20Interconnect%20Application%20Performance%20Brief.pdf},
  note         = {Accessed: 2025-04-16}
}

@article{grattafiori2024llama,
  title={The llama 3 herd of models},
  author={Grattafiori, Aaron and Dubey, Abhimanyu and Jauhri, Abhinav and Pandey, Abhinav and Kadian, Abhishek and Al-Dahle, Ahmad and Letman, Aiesha and Mathur, Akhil and Schelten, Alan and Vaughan, Alex and others},
  journal={arXiv preprint arXiv:2407.21783},
  year={2024}
}

@inproceedings{wu2025mirage,
  title={Mirage: A $\{$Multi-Level$\}$ Superoptimizer for Tensor Programs},
  author={Wu, Mengdi and Cheng, Xinhao and Liu, Shengyu and Shi, Chunan and Ji, Jianan and Ao, Man Kit and Velliengiri, Praveen and Miao, Xupeng and Padon, Oded and Jia, Zhihao},
  booktitle={19th USENIX Symposium on Operating Systems Design and Implementation (OSDI 25)},
  pages={21--38},
  year={2025}
}

@article{zhang2025comet,
  title={Comet: Fine-grained computation-communication overlapping for mixture-of-experts},
  author={Zhang, Shulai and Zheng, Ningxin and Lin, Haibin and Jiang, Ziheng and Bao, Wenlei and Jiang, Chengquan and Hou, Qi and Cui, Weihao and Zheng, Size and Chang, Li-Wen and others},
  journal={arXiv preprint arXiv:2502.19811},
  year={2025}
}

@inproceedings{autotvm,
author = {Chen, Tianqi and Zheng, Lianmin and Yan, Eddie and Jiang, Ziheng and Moreau, Thierry and Ceze, Luis and Guestrin, Carlos and Krishnamurthy, Arvind},
title = {Learning to optimize tensor programs},
year = {2018},
publisher = {Curran Associates Inc.},
address = {Red Hook, NY, USA},
booktitle = {Proceedings of the 32nd International Conference on Neural Information Processing Systems},
pages = {3393–3404},
numpages = {12},
location = {Montr\'{e}al, Canada},
series = {NIPS'18}
}

@inproceedings{ansor,
author = {Zheng, Lianmin and Jia, Chengfan and Sun, Minmin and Wu, Zhao and Yu, Cody Hao and Haj-Ali, Ameer and Wang, Yida and Yang, Jun and Zhuo, Danyang and Sen, Koushik and Gonzalez, Joseph E. and Stoica, Ion},
title = {Ansor: generating high-performance tensor programs for deep learning},
year = {2020},
isbn = {978-1-939133-19-9},
publisher = {USENIX Association},
address = {USA},
booktitle = {Proceedings of the 14th USENIX Conference on Operating Systems Design and Implementation},
articleno = {49},
numpages = {17},
series = {OSDI'20}
}

@article{shao2022tensor,
  title={Tensor program optimization with probabilistic programs},
  author={Shao, Junru and Zhou, Xiyou and Feng, Siyuan and Hou, Bohan and Lai, Ruihang and Jin, Hongyi and Lin, Wuwei and Masuda, Masahiro and Yu, Cody Hao and Chen, Tianqi},
  journal={Advances in Neural Information Processing Systems},
  volume={35},
  pages={35783--35796},
  year={2022}
}

@inproceedings{PrimePar,
author = {Wang, Haoran and Wang, Lei and Xu, Haobo and Wang, Ying and Li, Yuming and Han, Yinhe},
title = {PrimePar: Efficient Spatial-temporal Tensor Partitioning for Large Transformer Model Training},
year = {2024},
isbn = {9798400703867},
publisher = {Association for Computing Machinery},
address = {New York, NY, USA},
url = {https://doi.org/10.1145/3620666.3651357},
doi = {10.1145/3620666.3651357},
booktitle = {Proceedings of the 29th ACM International Conference on Architectural Support for Programming Languages and Operating Systems, Volume 3},
pages = {801–817},
numpages = {17},
location = {La Jolla, CA, USA},
series = {ASPLOS '24}
}

@article{zheng2025triton,
  title={Triton-distributed: Programming Overlapping Kernels on Distributed AI Systems with the Triton Compiler},
  author={Zheng, Size and Bao, Wenlei and Hou, Qi and Zheng, Xuegui and Fang, Jin and Huang, Chenhui and Li, Tianqi and Duanmu, Haojie and Chen, Renze and Xu, Ruifan and others},
  journal={arXiv preprint arXiv:2504.19442},
  year={2025}
}

@misc{nvidia_cute_dsl,
  author = {{NVIDIA}},
  title = {{CuTe DSL}},
  howpublished = {\url{https://docs.nvidia.com/cutlass/latest/media/docs/pythonDSL/cute_dsl.html}},
  year = {2026},
  note = {Accessed: 2026-04-18}
}

@misc{nvidia_cutile,
  author = {{NVIDIA}},
  title = {{cuTile Python}},
  howpublished = {\url{https://docs.nvidia.com/cuda/cutile-python/}},
  year = {2026},
  note = {Accessed: 2026-04-18}
}

@article{zheng2025tilelink,
  title={Tilelink: Generating efficient compute-communication overlapping kernels using tile-centric primitives},
  author={Zheng, Size and Fang, Jin and Zheng, Xuegui and Hou, Qi and Bao, Wenlei and Zheng, Ningxin and Jiang, Ziheng and Wang, Dongyang and Ye, Jianxi and Lin, Haibin and others},
  journal={arXiv preprint arXiv:2503.20313},
  year={2025}
}

@article{hong2025flashoverlap,
  title={FlashOverlap: A Lightweight Design for Efficiently Overlapping Communication and Computation},
  author={Hong, Ke and Li, Xiuhong and Liu, Minxu and Mao, Qiuli and Wu, Tianqi and Huang, Zixiao and Chen, Lufang and Wang, Zhong and Zhang, Yichong and Zhu, Zhenhua and others},
  journal={arXiv preprint arXiv:2504.19519},
  year={2025}
}

@techreport{NVIDIA_Hopper,
  author = {{NVIDIA}},
  title = {{NVIDIA H100 Tensor Core GPU Architecture}},
  institution = {{NVIDIA}},
  year = {2022},
  month = {mar},
  note = {White paper},
  url = {https://www.nvidia.com/content/dam/en-zz/Solutions/gtc22/data-center/h100/gtc22-whitepaper-hopper.pdf},
}

@inproceedings{mercury,
author = {Guan, Yue and Qiang, Xinwei and Pan, Zaifeng and Johnson, Daniels and Fang, Yuanwei and Zhou, Keren and Wang, Yuke and Li, Wanlu and Ding, Yufei and Aziz, Adnan},
title = {Mercury: Unlocking Multi-GPU Operator Optimization for LLMs via Remote Memory Scheduling},
year = {2025},
isbn = {9798400718700},
publisher = {Association for Computing Machinery},
address = {New York, NY, USA},
url = {https://doi.org/10.1145/3731569.3764798},
doi = {10.1145/3731569.3764798},
booktitle = {Proceedings of the ACM SIGOPS 31st Symposium on Operating Systems Principles},
pages = {1046–1061},
numpages = {16},
location = {Lotte Hotel World, Seoul, Republic of Korea},
series = {SOSP '25}
}

@misc{thunderkitten,
      title={ThunderKittens: Simple, Fast, and Adorable AI Kernels}, 
      author={Benjamin F. Spector and Simran Arora and Aaryan Singhal and Daniel Y. Fu and Christopher Ré},
      year={2024},
      eprint={2410.20399},
      archivePrefix={arXiv},
      primaryClass={cs.LG},
      url={https://arxiv.org/abs/2410.20399}, 
}

@misc{parallelkitten,
      title={ParallelKittens: Systematic and Practical Simplification of Multi-GPU AI Kernels}, 
      author={Stuart H. Sul and Simran Arora and Benjamin F. Spector and Christopher Ré},
      year={2025},
      eprint={2511.13940},
      archivePrefix={arXiv},
      primaryClass={cs.DC},
      url={https://arxiv.org/abs/2511.13940}, 
}

@misc{wang2024dominoeliminatingcommunicationllm,
      title={Domino: Eliminating Communication in LLM Training via Generic Tensor Slicing and Overlapping}, 
      author={Guanhua Wang and Chengming Zhang and Zheyu Shen and Ang Li and Olatunji Ruwase},
      year={2024},
      eprint={2409.15241},
      archivePrefix={arXiv},
      primaryClass={cs.DC},
      url={https://arxiv.org/abs/2409.15241}, 
}

@online{nvidia_sharp_v300,
  author       = {NVIDIA},
  title        = {{NVIDIA Scalable Hierarchical Aggregation and Reduction Protocol (SHARP) Rev 3.0.0}},
  year         = {2024},
  url          = {https://docs.nvidia.com/networking/display/sharpv300},
  note         = {Accessed: 2025-02-10}
}

@misc{zhu2025nanoflowoptimallargelanguage,
      title={NanoFlow: Towards Optimal Large Language Model Serving Throughput}, 
      author={Kan Zhu and Yufei Gao and Yilong Zhao and Liangyu Zhao and Gefei Zuo and Yile Gu and Dedong Xie and Tian Tang and Qinyu Xu and Zihao Ye and Keisuke Kamahori and Chien-Yu Lin and Ziren Wang and Stephanie Wang and Arvind Krishnamurthy and Baris Kasikci},
      year={2025},
      eprint={2408.12757},
      archivePrefix={arXiv},
      primaryClass={cs.DC},
      url={https://arxiv.org/abs/2408.12757}, 
}

@misc{gond2025tokenweaveefficientcomputecommunicationoverlap,
      title={TokenWeave: Efficient Compute-Communication Overlap for Distributed LLM Inference}, 
      author={Raja Gond and Nipun Kwatra and Ramachandran Ramjee},
      year={2025},
      eprint={2505.11329},
      archivePrefix={arXiv},
      primaryClass={cs.DC},
      url={https://arxiv.org/abs/2505.11329}, 
}

@inproceedings{OverlapCommunicationwithDependentComputationviaDecomposition,
author = {Wang, Shibo and Wei, Jinliang and Sabne, Amit and Davis, Andy and Ilbeyi, Berkin and Hechtman, Blake and Chen, Dehao and Murthy, Karthik Srinivasa and Maggioni, Marcello and Zhang, Qiao and Kumar, Sameer and Guo, Tongfei and Xu, Yuanzhong and Zhou, Zongwei},
title = {Overlap Communication with Dependent Computation via Decomposition in Large Deep Learning Models},
year = {2022},
isbn = {9781450399159},
publisher = {Association for Computing Machinery},
address = {New York, NY, USA},
url = {https://doi.org/10.1145/3567955.3567959},
doi = {10.1145/3567955.3567959},
booktitle = {Proceedings of the 28th ACM International Conference on Architectural Support for Programming Languages and Operating Systems, Volume 1},
pages = {93–106},
numpages = {14},
location = {Vancouver, BC, Canada},
series = {ASPLOS 2023}
}

@misc{qwen2025qwen25technicalreport,
      title={Qwen2.5 Technical Report}, 
      author={Qwen and : and An Yang and Baosong Yang and Beichen Zhang and Binyuan Hui and Bo Zheng and Bowen Yu and Chengyuan Li and Dayiheng Liu and Fei Huang and Haoran Wei and Huan Lin and Jian Yang and Jianhong Tu and Jianwei Zhang and Jianxin Yang and Jiaxi Yang and Jingren Zhou and Junyang Lin and Kai Dang and Keming Lu and Keqin Bao and Kexin Yang and Le Yu and Mei Li and Mingfeng Xue and Pei Zhang and Qin Zhu and Rui Men and Runji Lin and Tianhao Li and Tianyi Tang and Tingyu Xia and Xingzhang Ren and Xuancheng Ren and Yang Fan and Yang Su and Yichang Zhang and Yu Wan and Yuqiong Liu and Zeyu Cui and Zhenru Zhang and Zihan Qiu},
      year={2025},
      eprint={2412.15115},
      archivePrefix={arXiv},
      primaryClass={cs.CL},
      url={https://arxiv.org/abs/2412.15115}, 
}

@inproceedings{tacos,
author = {Won, William and Elavazhagan, Midhilesh and Srinivasan, Sudarshan and Gupta, Swati and Krishna, Tushar},
title = {TACOS: Topology-Aware Collective Algorithm Synthesizer for Distributed Machine Learning},
year = {2024},
publisher = {IEEE Press},
url = {https://doi.org/10.1109/MICRO61859.2024.00068},
doi = {10.1109/MICRO61859.2024.00068},
booktitle = {Proceedings of the 2024 57th IEEE/ACM International Symposium on Microarchitecture},
pages = {856–870},
numpages = {15},
location = {Austin, TX, USA},
series = {MICRO '24}
}

@misc{openmp,
  title        = {OpenMP Application Programming Interface},
  author       = {{OpenMP Architecture Review Board}},
  year         = {2023},
  howpublished = {\url{https://www.openmp.org/specifications/}}
}

@misc{xu2021gspmdgeneralscalableparallelization,
      title={GSPMD: General and Scalable Parallelization for ML Computation Graphs}, 
      author={Yuanzhong Xu and HyoukJoong Lee and Dehao Chen and Blake Hechtman and Yanping Huang and Rahul Joshi and Maxim Krikun and Dmitry Lepikhin and Andy Ly and Marcello Maggioni and Ruoming Pang and Noam Shazeer and Shibo Wang and Tao Wang and Yonghui Wu and Zhifeng Chen},
      year={2021},
      eprint={2105.04663},
      archivePrefix={arXiv},
      primaryClass={cs.DC},
      url={https://arxiv.org/abs/2105.04663}, 
}

@misc{shazeer2018meshtensorflowdeeplearningsupercomputers,
      title={Mesh-TensorFlow: Deep Learning for Supercomputers}, 
      author={Noam Shazeer and Youlong Cheng and Niki Parmar and Dustin Tran and Ashish Vaswani and Penporn Koanantakool and Peter Hawkins and HyoukJoong Lee and Mingsheng Hong and Cliff Young and Ryan Sepassi and Blake Hechtman},
      year={2018},
      eprint={1811.02084},
      archivePrefix={arXiv},
      primaryClass={cs.LG},
      url={https://arxiv.org/abs/1811.02084}, 
}

\appendix
\clearpage
\section{Artifact Appendix}
\label{sec:artifact}

\subsection*{Abstract}

The \thiswork{} artifact contains the source code for the chunk-centric
compiler and runtime described in this paper, together with unit tests,
end-to-end multi-GPU tests, experiment entry points, and a Dockerfile for the
GPU software environment. A short CPU-only workflow validates the core
communication-plan representation, signal planning, lowering logic, and
compiler passes. On a multi-GPU NVIDIA Hopper-class system, the artifact also
supports correctness and performance evaluation of generated GEMM and
attention operators.

\subsection*{Scope}

The artifact can be used to validate three central claims of the paper:
(1) chunk-level communication plans can express reusable communication
schedules and lower them to per-rank execution plans; (2) \thiswork{} can
transform local tiled kernels into correct fine-grained overlapped multi-GPU
kernels; and (3) the generated operators can be benchmarked across
communication backends and scheduling choices. The CPU-only tests validate the
compiler abstractions and passes, while the GPU tests validate generated-kernel
correctness against unfused references and report execution time.

Reproducing all absolute performance numbers and comparisons in the paper
requires a 4- or 8-GPU Hopper server with NVLink/NVSwitch connectivity and the
external baseline implementations. Several external baselines are not vendored
in the artifact; the repository documents this limitation and points to their
upstream repositories.

\subsection*{Contents}

The repository is organized as follows:
\begin{itemize}
    \item \texttt{syncopate/communication/}: communication-plan descriptors,
    signal planning, code generation, and runtime support;
    \item \texttt{syncopate/computation/}: GEMM and attention kernel templates
    and the source-to-source transformation pass;
    \item \texttt{syncopate/interface/}: lowering from communication plans to
    per-rank schedules and tile schedules;
    \item \texttt{tests/}: CPU-only compiler tests and end-to-end multi-GPU
    correctness/performance tests;
    \item \texttt{experiments/}: figure-oriented experiment entry points for
    the operator, attention, compiler-integration, and ablation results; and
    \item \texttt{utils/}: the distributed launcher, communication
    microbenchmarks, and helper scripts.
\end{itemize}

\subsection*{Hosting}

The artifact is publicly available under the MIT license at
\url{https://github.com/tie-pilot-qxw/syncopate}. The version evaluated for
OSDI~'26 Artifact Evaluation is commit
\texttt{ce2b13e496eb6629ab42aac2aec1d9e31e084ba6}. Evaluators and readers
should check out this commit to obtain the fixed artifact version.

\subsection*{Requirements}

The CPU-only validation requires Linux, Python~3.10 or newer, PyTorch, and
\texttt{pytest}; it typically completes in under five minutes. Full
generated-kernel and performance experiments require Linux, Docker with the
NVIDIA Container Toolkit, and a single node with at least four
NVLink/NVSwitch-connected NVIDIA Hopper-class GPUs; some experiments require
eight GPUs.

%%%%%%%%%%%%%%%%%%%%%%%%%%%%%%%%%%%%%%%%%%%%%%%%%%%%%%%%%%%%%%%%%%%%%%%%%%%%%%%%
\end{document}